\documentclass[11pt]{article}
\usepackage[utf8]{inputenc}
\usepackage{authblk}
\usepackage{natbib}
\usepackage{graphicx}
\usepackage{adjustbox}
\usepackage{geometry}
\usepackage{caption}
\usepackage{subcaption}
\usepackage{hyperref}
\hypersetup{
    colorlinks,
    linkcolor={red!50!black},
    citecolor={blue!50!black},
    urlcolor={blue!80!black}
}
\usepackage{amsmath,amssymb}
\usepackage{bm}
\usepackage{multicol}
\usepackage{multirow}
\usepackage{booktabs}
\usepackage{comment}
\usepackage{xcolor}
\usepackage{soul}
\usepackage{tikz}
\usepackage{tikz-network}
\usepackage{pgfplots}
\usepackage{array}

\newcommand{\indep}{\perp \!\!\! \perp}
\newcommand{\xdownarrow}[1]{%
  {\left\downarrow\vbox to #1{}\right.\kern-\nulldelimiterspace}
}

\title{Cultures as networks of cultural traits:\\ {\large A unifying framework for measuring culture and cultural distances}.}
\author[1,4]{Luca De Benedictis}
\author[2]{Roberto Rondinelli\thanks{\scriptsize \ The authors would like to acknowledge the contribution of the COST Action CA15109 (COSTNET), which partially funded a visit of Roberto Rondinelli to Brunel University in London. Many thanks to Ernst Wit, to the participants to various seminars and conferences (SIE2021, EUSN2021, \textcolor{black}{SUNBELT2022, EGI2022}),
\textcolor{black}{and to the anonymous referees, the Associate Editor and the Joint Editor of the journal for the insightful comments}.  
No conflict of interest applies.}}
\author[3]{Veronica Vinciotti}
\affil[1]{\small University of Macerata - luca.debenedictis@unimc.it}
\affil[2]{\small University of Naples Federico II - roberto.rondinelli@unina.it}
\affil[3]{\small University of Trento - veronica.vinciotti@unitn.it}
\affil[4]{\small Luiss University}
\date{}

\begin{document}
\sethlcolor{lime} 

\maketitle

\begin{abstract}
\footnotesize{ 
Making use of the information from the World Value Survey (WVS), and operationalizing a definition of national culture that encompasses both the relevance of specific cultural traits \emph{and} the interdependence among them, this paper proposes a methodology to reveal the latent structure of national culture and to measure cultural distance between countries that takes into account both the difference in cultural traits and the difference in the network structure of national cultures.
Exploiting the possibilities offered by copula graphical models for discrete data, this paper infers the 
\textcolor{black}{cultural networks of all the countries included in the WVS (Wave 6)} and proposes a novel unifying framework to measure national culture and international cultural distances. The Jeffreys' divergence between copula graphical models, 
taken as the measure of cultural distance between countries, captures the orthogonality of the two components of cultural distance: the one based on cultural traits and the one based on the network structure among them. Moreover, the two components are shown to correlate with different national and structural characteristics of cultural networks, thus encompassing the different informational sets related to national cultures. 
}
\end{abstract}

\vspace{2mm}


\footnotesize
\vspace{2mm}

\section{Introduction}
\label{intro}

The notion of culture - in the broad sense of local norms, customs, attitudes, values, 
and their subsets \emph{and interactions}
-- pervades societies and the way they are studied in the social sciences (see \cite{cuche2020notion} for a recent overview, \cite{taylor1871primitive} for an early account in cultural anthropology, and \cite{united2001universal} for an institutional definition). Recently, a large amount of studies have been focused on the quantification of culture and its interplay with political issues \citep{lane2016culture}, economic growth \citep{tabellini2008institutions, AleGiu2015}, comparative sociology \citep{schwartz2008cultural}, management \citep{yeganeh2006conceptual}, anthropology \citep{ruck2020cultural}, and psychology \citep{kashima2019psychology}.
However, two fundamental issues lie at the basis of empirical studies on culture: a consensual \emph{definition} of the object of analysis and a well defined \emph{measurement} of it, \textcolor{black}{that could emphasise the interconnected dimension of culture}.
\newline
\newline
As far as the first issue is concerned, from \cite{taylor1871primitive} on, the definition of culture has evolved along time, offering many possible alternative or compatible specifications. \cite{Kro-Klu1952} recorded 160 possible definitions of culture,
and this list is by far exhaustive. Revisions of the relevant \emph{cultural traits}
- being them, citing the classification of \cite{huxley1880coming}, mentifacts (e.g. ideas, values, and beliefs), sociofacts (e.g social structures) or artifacts (e.g. goods \citep{boas1982race} and technologies) - are periodically proposed by different scholars from different disciplines
\citep{dawkins2016selfish}. Moreover, the definition itself can be mediated by the geographical origin, the historical context, the personal experience and the specific scope of the scholar proposing it,
so, quite often, comparative analyses of culture conclude that, because of the specificity of national cultures, there \emph{cannot be} any generally agreed definition of culture  \citep{Jho2012}.  
\newline
\newline
Regarding measurement, cultures and cultural traits have been measured in many different ways, from field work, collecting objects or individual opinions and recording evidences of cultural heritages, to lab or on-the-field experiments, in which people play Trust, Public good, Dictator or Ultimatum games \citep{RotPraOkuZam1991,OosSloVan2004}. \textcolor{black}{Recently, the growing interest in cross-country cultural studies, and the requirement of systematic and comparative data, have moved the attention to information collected through social surveys\textcolor{black}{, in the tradition of \cite{hofstede1984culture}'s seminal research project (see also \cite{schwartz1994beyond} for a subsequent influential research, and \cite{taras2009half} for a recent overview of the issue)}}. Some of the surveys are multi-country surveys, such as the Life in Transition Survey (LITS) 
or the one used in the present study, the World Values Survey (WVS). Others are at the regional level, focusing on one geographical or political area (e.g. the European Values Study or the Eurobarometer, for the EU \citep{GuiSapZin2009}),
or at the national level,  focusing on different spatial units of one single country (e.g. the General Social Survey (GSS), for the US \citep{AleLaF2000, AleGiu2011, GiuSpi2014}). 
\textcolor{black}{In all cases, the surveys report the opinion of a sample of the population on individual values, attitudes, customs or local norms, that can reflect cultural traits. The aggregation of individual opinions and the prevalence of some cultural traits is then used to define national cultures, whereas cross-country comparisons are used to measure the cultural distance between countries or cluster of countries.}
\newline
\newline
This paper shares the same objective of many researches on cross-country comparative cultural analyses: defining and quantifying the notion of national culture and measuring the cultural distance between countries. It builds on 
the many contributions of \cite{inglehart2015silent,inglehart2018culture,inglehart1997modernization,Ing2018}, and, by making use of existing data from Wave 6 (2010-2014) of the WVS \citep{Ing2014}, it quantifies the relevance of cultural traits for every single country included in the original Inglehart and Welzel's index of cultural distance \citep{IngWel2005}, named \texttt{IW index} of cultural distance, for short. But from this point on, the approach proposed in this paper offers a new perspective on how cultural traits define national culture and how to measure the cultural distance between countries. In fact, 
the view that this paper offers departs from the one of other studies on cross-country comparative cultural analyses, by assuming that, for \emph{any} finite set of cultural traits, the resulting national culture is more that the sum of its parts. Indeed, national cultures are made of relevant cultural traits \emph{and} of the latent network structure among them. Analogously, the cultural distance between countries is not only dependent on the relative relevance of single cultural traits but also on the \emph{interdependence} among common cultural traits that characterises any specific national culture.
\newline
\newline
A simple example illustrates the main idea.
Imagine there are two countries, and two cultural traits, with each trait taking two possible values (e.g.: belief in God
(yes/no) and trust in others (yes/no)). Imagine that in both countries, half the people believe in God and
half the people trust others. The two countries would appear to be at a cultural distance of zero. However,
suppose that in the first country everyone who believes in God also trusts others, while in the second country everyone who believes in God does not trust others. Then, the two countries are actually culturally different
because the pattern of interdependence between cultural traits across individuals within each country is
different. The higher the number of cultural traits the stronger the potential influence of the interaction among them. Consequently, measuring the cultural distance between countries without taking into account the network structure of cultural traits would result in a systematic bias of potentially relevant magnitude.
\newline
\newline
\textcolor{black}{This new view of culture, based on the superadditivity of cultural traits, exploits the possibilities offered by graphical models \citep{lauritzen1996graphical} to uncover the latent structure of cultural traits' interdependence.} Under the recently developed Bayesian inferential scheme for discrete data \citep{mohammadi2017bayesian}, first, it quantifies national cultures as networks of relevant cultural traits, and, second, it measures the distance between national cultures considering not only the significance of cultural traits per sè, but also the resulting \emph{interdependence} among cultural traits. This approach results in a new index of cultural distance, a $f$-divergence measure that takes the form of a Jeffreys' divergence, named \texttt{JD index} of cultural distance. This distance is naturally defined as the sum of two orthogonal components, one depending on cultural traits (\texttt{JD marginals}) and the other depending on the interdependence among them (\texttt{JD network}). While  \texttt{JD marginals} has a high correlation with the original \texttt{IW index} (Pearson's correlation value of 0.89), the \texttt{JD network} is uncorrelated with it (Pearson's correlation value of -0.03) and captures countries' differences at the network level.
\newline
\newline 
The analysis presented in this paper shows how the \texttt{JD index} allows to measure cultural distance in a unified framework, considering orthogonal information and showing a dimension of cultural heterogeneity largely unexplored by the literature on comparative cultural analyses, until now. Moreover, from an applied point of view, an adequate measurement and representation of the cultural heterogeneity among countries is essential in providing adequate information for the academic and the public debate about the uniqueness, similarities and diversity of national societies.
\newline
\newline
The paper is organised as follow. Section \ref{WVS} includes a concise description of the WVS data, of the cultural traits included in the \texttt{IW index}, and of the original Cultural Map derived from it.  
Sections \ref{methods} is dedicated to the description of the proposed methodology for network inference, based on copula graphical models, and to the \texttt{JD index} of cultural distance that is derived from that. Section \ref{sec:results} starts with an analysis of the national cultural networks inferred from the WVS, it then focuses on their main topological components and thus proceeds with an evaluation of their distances using the proposed index. 
Finally, performing dyadic regressions of \texttt{JD marginals} and \texttt{JD network} on several different explanatory variables, it reveals the main elements determining the cultural distances between countries.
Section \ref{conclusions} concludes with some possible future extensions of the analysis to different sets of cultural traits, to a different pool of countries, to different country groups, and to the repeated cross-sectional data derived from multiple surveys.

\section{World Values Survey}
\label{WVS}
    
The WVS is a cross-country research project carried out for almost 30 years. The resulting database is public and freely accessible online, making it the most widely used social survey database in the world, and making the \cite{IngWel2005} Cultural Map one of the most cited and used tool of analysis in cross-cultural studies. This, and the \texttt{IW index} associated to it, is going to be the benchmark of the present study.
\newline
\newline
The original purpose of the WVS was to test the idea that economic conditions are changing the fundamental attitudes and values in industrialised countries. In doing so, the researchers designed questionnaires
inquiring respondents about religion, political preferences, attitudes and values. From the original intent, the project expanded rapidly 
to a true global investigation on world cultural traits.
\newline
\newline
Since 1977 the WVS has completed six waves of polls. WVS Wave 1 (1981-1984) covered 24 countries and, with a questionnaire of 268 questions, presented evidences for inter-generational shifts in cultural traits \citep{inglehart2018culture}. 
WVS Wave 6 (2010-2014) is now including 60 countries with more than 85,000 respondents completing the questionnaire through face-to-face or phone interviews.  In each country \textcolor{black}{around 1000 individuals are interviewed (with a minimum of 841 for New Zealand and a maximum of 4078 for India)} and samples are representative of the national adult population (18 years and older) \citep{Ing2014}. The questionnaire is composed of 258 questions, whose answers are mainly ordinal or nominal. 
In every country the questions are translated and accurately adapted to national specificity.
The resulting data are the ones used in the present analysis.

\subsection{IW Index cultural traits}
\label{data}
        
In their construction of the Cultural Map, \cite{IngWel2005}  use a subset of the questions included in the WVS  questionnaire and derive a measure of national culture from 10 cultural traits.
In order to be consistent with the original setup, the same selection of questions used to construct specific variables defining the cultural traits, the same countries and the same country groups used in the original Cultural Map are considered also in this paper. The selection of a different set of cultural traits is of course possible, and the methodology proposed in the paper can be applied to different data with similar characteristics.
\newline
\newline
Referring to WVS Wave 6 \citep{Ing2014}, the 10 cultural traits have been quantified in the 10 corresponding variables, identified by a different font: 
\texttt{happiness}, (V10), quantifies the self reported level of happiness in an ordered scale from 1 (``Not at all happy'') to 4 (``Very happy''); \texttt{trust}, (V24), measures trust in others as a dichotomy, with 1 indicating that ``Most people can be trusted'' and 2 indicating that ``You have to be careful.'' \texttt{respect for authority}, (V69), records the approval for a general greater respect for authority, on an ordered scale from 1 (``is a good thing'') to 3 (``is a bad thing''); while \texttt{voice}, (V85), in the sense of \cite{Hir1970}, measures the availability of people to express personal opinions by signing petitions, in an ordered scale from 1 (``did sign'') to 3 (``would never do'').
The \texttt{importance of God} in a personal life, (V152), justification of \texttt{homosexuality}, (V203),  or \texttt{abortion}, (V204), record personal values and opinions in terms of religiosity and morality on a scale from 1 (``not at all important,'' referring to God; ``never be justified,'' referring to homosexuality and abortion) to 10 (``very important,'' referring to God; ``always justified,'' referring to homosexuality and abortion). 
\texttt{national pride}, (V211), quantifies the pride to be a citizen of a certain country, in an ordered scale from 1 (``very proud'') to 4 (``not at all proud'').
\texttt{post-materialism}, in sociology, is the transformation of individual values from materialist, physical, and economic to progressive  individual values of autonomy and self-expression \citep{Ing2018}. 
 Here the variable (Y002) is a composite index associated to individual rankings of social values, such as the preference between order, freedom, economic stability, and public participation, and it is coded with 1 indicating Materialism, 2 Mixed, and 3 Post-materialism. Finally, \texttt{obedience/independence}, (Y003), measures the importance of teaching children to have faith in God, to obey, to be independent, to pursue perseverance and determination, on a scale from 1 (``obedience'') to 5 (``independence''). The original coding has been maintained for consistency with \citet{IngWel2005}.

\begin{table}[!ht]
\begin{center}
\caption{\small{\textit{Descriptive statistics}}}
\begin{scriptsize}
\begin{tabular}{lllccc}
\toprule
\multicolumn{3}{c}{Cultural traits} & & & \multicolumn{1}{c}{} \\
\cmidrule(l){1-3} 
id    & \texttt{label} & \texttt{variable} (categories)     & distribution      &  overall mean      & missing \%\\
    &  &      &       &  [min, max]       & [min, max] \\
\midrule
\multirow{2}*{V10} & \multirow{2}*{\texttt{H}} &\multirow{2}*{level of \texttt{happiness} (1:4)}  & \includegraphics[width=0.075\linewidth]{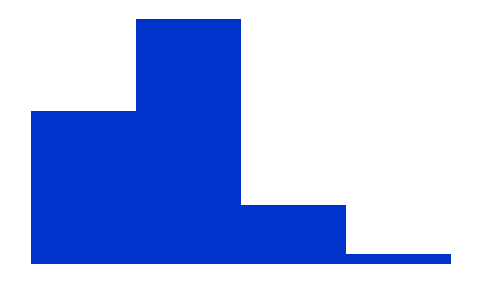}        & 1.855           & 0.8          \\[-0.5ex]
                &        & & [{\tiny low \  \ high}]  & [1.387, 2.256]           & [0.0, 5.5]    \\[0.5ex]

\multirow{2}*{V24} & \multirow{2}*{\texttt{T}}& \multirow{2}*{\texttt{trust} in people  (1:2)}  &  \includegraphics[width=0.075\linewidth]{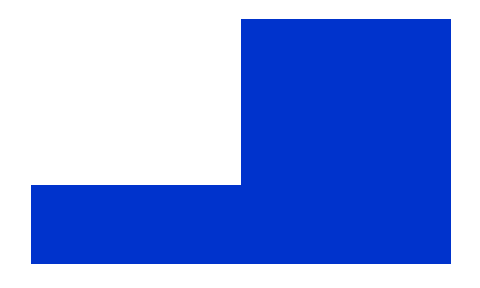}          & 1.764       & 2.7            \\[-0.5ex]
                & &          & [{\tiny high \  \ low}]           & [1.326, 1.971]   & [0.00, 10.8]    \\[0.5ex]

\multirow{2}*{V69} & \multirow{2}*{\texttt{R}}& \multirow{2}*{\texttt{respect for authority} (1:3) }   &  \includegraphics[width=0.075\linewidth]{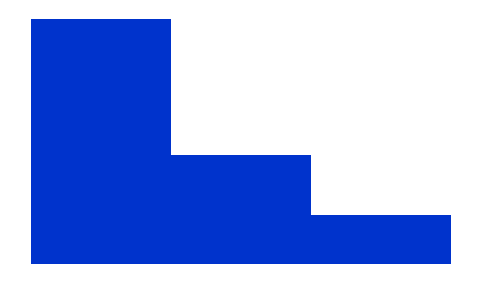}        &  1.503          & 4.5                 \\[-0.5ex]
                & &                  & [{\tiny high \  \ low}]          & [1.062, 2.732]          & [0.00, 18.7]                  \\[0.5ex]

\multirow{2}*{V85} & \multirow{2}*{\texttt{V}} & \multirow{2}*{\texttt{voice} through petitions (1:3) }  & \includegraphics[width=0.075\linewidth]{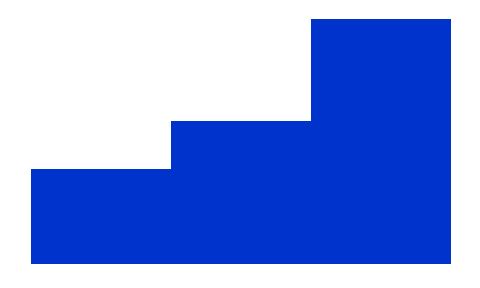}        & 2.314            & 5.2         \\[-0.5ex]
                & &        & [{\tiny high \  \ low}]  &  [1.19, 2.857]          & [0.00, 30.2]    \\[0.5ex]

\multirow{2}*{V152} & \multirow{2}*{\texttt{G}} & \multirow{2}*{\texttt{importance of God} (1:10) }  &  \includegraphics[width=0.075\linewidth]{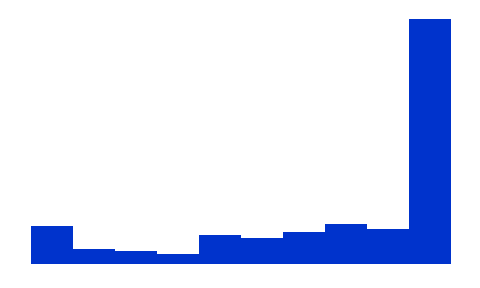}          & 7.781       & 2.5            \\[-0.5ex]
                & &          &[{\tiny low \  \ high}]           & [3.482, 9.865]   & [0.00, 16.3]    \\[0.5ex]

\multirow{2}*{V203} & \multirow{2}*{\texttt{O}} & \multirow{2}*{justification of \texttt{homosexuality} (1:10) }   &  \includegraphics[width=0.075\linewidth]{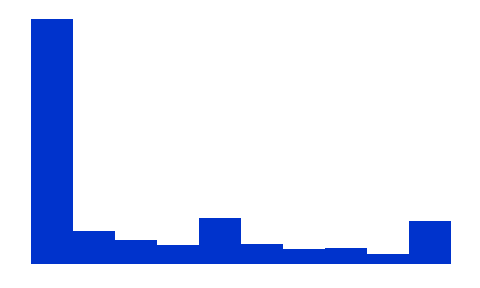}        & 3.35            &  5.6                \\[-0.5ex]
                & &                  & [{\tiny low \  \ high}]          & [1.127, 8.47]           & [0.00, 34.6]                \\[0.5ex]

\multirow{2}*{V204} & \multirow{2}*{\texttt{A}} & \multirow{2}*{justification of \texttt{abortion} (1:10)}  & \includegraphics[width=0.075\linewidth]{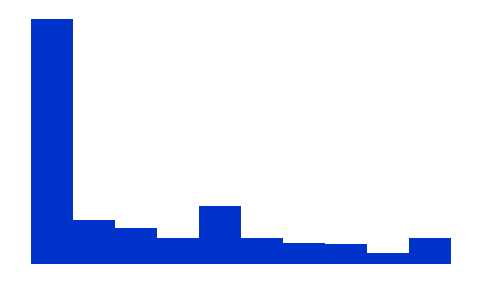}        & 3.24            & 3.8          \\[-0.5ex]
                & &         & [{\tiny low \  \ high}]  & [1.515, 7.997]            & [0.00, 19.0]    \\[0.5ex]

\multirow{2}*{V211} & \multirow{2}*{\texttt{P}} & \multirow{2}*{\texttt{national pride} (1:4)}  &   \includegraphics[width=0.075\linewidth]{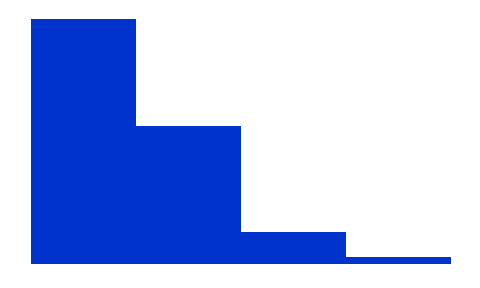}         & 1.523       & 2.8           \\[-0.5ex]
                & &           & [{\tiny high \  \ low}]          & [1.059, 2.176]   & [0.00, 12.3]    \\[0.5ex]

\multirow{2}*{Y002} & \multirow{2}*{\texttt{M}} & \multirow{2}*{\texttt{post-materialism} (1:3)}   &  \includegraphics[width=0.075\linewidth]{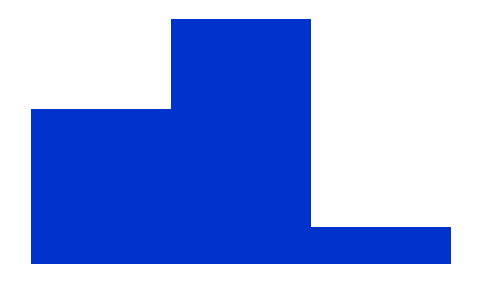}        & 1.73           & 4.2                 \\[-0.5ex]
                & &                   & [{\tiny low \  \ high}]          & [1.281, 2.244]         & [0.00, 31.9]                \\[0.5ex]

\multirow{2}*{Y003} & \multirow{2}*{\texttt{B}} & \multirow{2}*{\texttt{obedience vs independence}   (1:5)}   &  \includegraphics[width=0.075\linewidth]{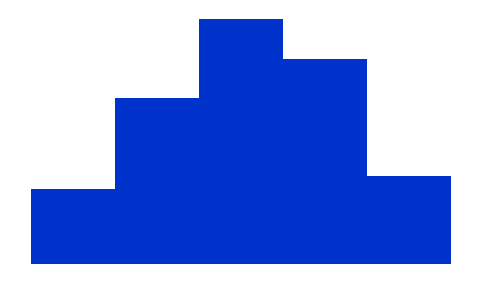}        &  3.059          &  0.3                \\[-0.5ex]
                & &                 &  [{\tiny high \  \ low}]         & [2.066, 4.261]         & [0.00, 1.0]                 \\[0.5ex]
\bottomrule
\end{tabular}
\end{scriptsize}
\label{tab:descriptives}
\end{center}
{\scriptsize {\bf Note}: {\linespread{0.5}\selectfont Column id includes the WVS Wave 6 questionnaire codes of the 10 selected cultural traits originally by \cite{IngWel2005}. Column \texttt{label} contains the one-letter shortening of the variable, that is going to be used in Figure \ref{fig:nodedyadtriad} and \textcolor{black}{Tables \ref{tab:dyadtriad}, \ref{tab:BayesianCI} and \ref{tab:reg2}}; while \texttt{variable} contains a longer description of the variable, generally used in the main text, and specifies for the corresponding variable the number of ordinal categories (i.e. (1:4) means that the variable assumes 4 ordered categories). The histograms, included in the forth column, refer to the overall distributions of all responses in the 54 countries, for every variable; the two terms below each histogram indicate the direction of the categorical ordering. Column five and six contain the overall average value of every cultural trait and the percentage of missing values; and the [min,max] intervals refer to the lowest/highest values of mean and \% of missing, respectively, across the 54 countries. 
}
\par}
\end{table}

Table \ref{tab:descriptives}, that summarises the characteristics of the different cultural traits for all countries considered, shows three groups of variables as far as missing observations are concerned: missing are minimal for \texttt{obedience/inde- pendence} and \texttt{happiness}; around 2.5\% for \texttt{national pride} and \texttt{trust}; and around 4.5\% for all the other variables. More importantly, there is a high variability at the country-level for the responses regarding sensitive questions: indeed, missingness rises above 30\% for \texttt{voice} in the case of Algeria, \texttt{homosexuality} for Yemen, and \texttt{post-materialism} for New Zealand. China is the country with the highest number of missing values for \texttt{importance of God}, \texttt{abortion}, and \texttt{national pride}. While high levels of missingness could  introduce a selection bias in the reported values for several variables, which means that measures of cultural distance have to be taken with caution for the more extreme cases, it is worth emphasising how the methodology proposed in this paper accounts for missing values without the need for discarding the responses that are only partially completed, which is a common feature of survey data.
\newline
\newline
The mean value of the variables considered depends on the number of ordinal categories and can vary substantially from country to country. As an example, \texttt{happiness} shows an overall mean for all countries of 1.855, out of four categories coded from 1 to 4, with some countries being characterised by a low level of \texttt{happiness} (the minimum mean value of 1.387 corresponds to Mexico) while others record a much higher value (the maximum mean value of 2.256 corresponds to Iraq).  
\newline
\newline
The [min, max] intervals 
show a large heterogeneity across countries. The country recording the minimum value in the variable \texttt{trust} is The Netherlands, and the one with the maximum value is The Philippines. Since the variable is coded from trusting to not trusting, The Netherlands is a country in which the trust in people is high, while the reverse is true for the Philippines. 
The maximum in the average value of \texttt{respect for authority} corresponds instead to Japan, where - given the reverse coding of the variable - most of the people think that a further increase in the respect for authority would be a bad thing; the opposite extreme case is the one of Ghana where most people think that a further increase in the respect for authority would be a good thing. The same country records the highest average \texttt{national pride} (1.059), while the opposite case is represented by Taiwan. Expressing personal opinions through petitions is relevant in New Zealand, with an average of 1.19, while it is not an option in Azerbaijan, with an average of 2.857. 
China is the country where the \texttt{importance of God} in personal life is minimal (average of 3.482), while it is maximal in the case of Yemen. Similarly, \texttt{homosexuality} is almost never justified in Armenia (average 1.127), as \texttt{abortion} is in Pakistan (average 1.515); while the two are largely justified in Sweden, with averages of 8.47 and  7.997, respectively. Tunisia scores the lowest average level of \texttt{post-materialism} (1.281) and Sweden the highest one (2.243). Finally, \texttt{obedience} is important in Yemen, with an average of \texttt{B} equal to 2.066, while \texttt{independence} is important in Japan, with an average of 4.261.
\newline
\newline
Comparing the distributions of the variables with their mean highlights the limitation of existing methodologies that compare cultural traits by only focusing on the first moment of the distribution, while disregarding its higher moments. For example, \texttt{B} and \texttt{A} have very similar mean values, but the distribution of the former is close to symmetric while the one of the latter is strongly skewed to the left. Similarly, in the cases of \texttt{T}, \texttt{R} and the other asymmetric variables, the mean value is not a sufficiently representative statistic.

\subsection{\textcolor{black}{Reproducing} the Inglehart-Welzel Cultural Map}
\label{IWmap}
       
\begin{figure}[!ht]
            \begin{center}
                \caption{\small{\textit{Principal Component Analysis on Inglehart-Welzel Cultural Map variables.}}}
                \includegraphics[width=0.5\linewidth]{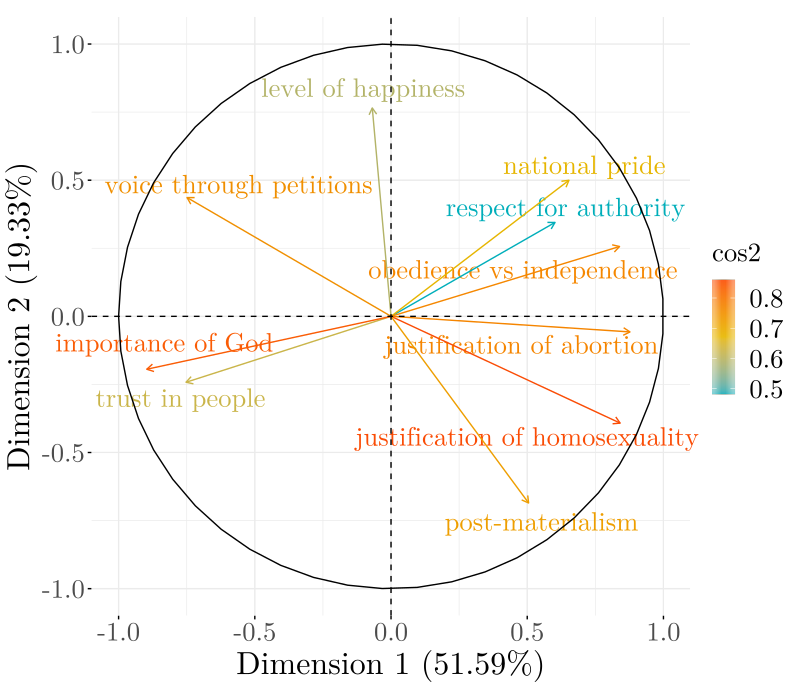} 

                \label{fig:pca}
            \end{center}
                 \vspace{-0.5cm}
    {\scriptsize {\bf Note}: {\linespread{0.5}\selectfont Authors' elaborations on WVS Wave 6. The variables in the correlation circle are:
V10 - level of \texttt{happiness}; 
V24 - \texttt{trust} in people; 
V69 - \texttt{respect for authority};
V85 - \texttt{voice} through petitions;
V152 - \texttt{importance of God};
V203 - justification of \texttt{homosexuality};
V204 - justification of \texttt{abortion};
V211 - \texttt{national pride};
Y002 - \texttt{post-materialism};
Y003 - \texttt{obedience vs independence}. Missing values are not considered. Recall from Table \ref{tab:descriptives} that \texttt{trust} in people (\texttt{T}), \texttt{respect for authority} (\texttt{R}), \texttt{voice} through petitions (\texttt{V}), \texttt{national pride} (\texttt{P}), and \texttt{obedience vs independence} (\texttt{B}) are coded in descending order from high to low. Dimension 1 projects the First principal component; Dimension 2 projects the Second principal component. Colours correspond to the product of the Squared cosine (cos2) of the two principal components. \par}}
\end{figure}

\begin{figure}[!ht]
                \begin{center}
                \caption{\small{\textit{A reproduction of Inglehart-Welzel Cultural Map}}}
                \includegraphics[width=\textwidth]{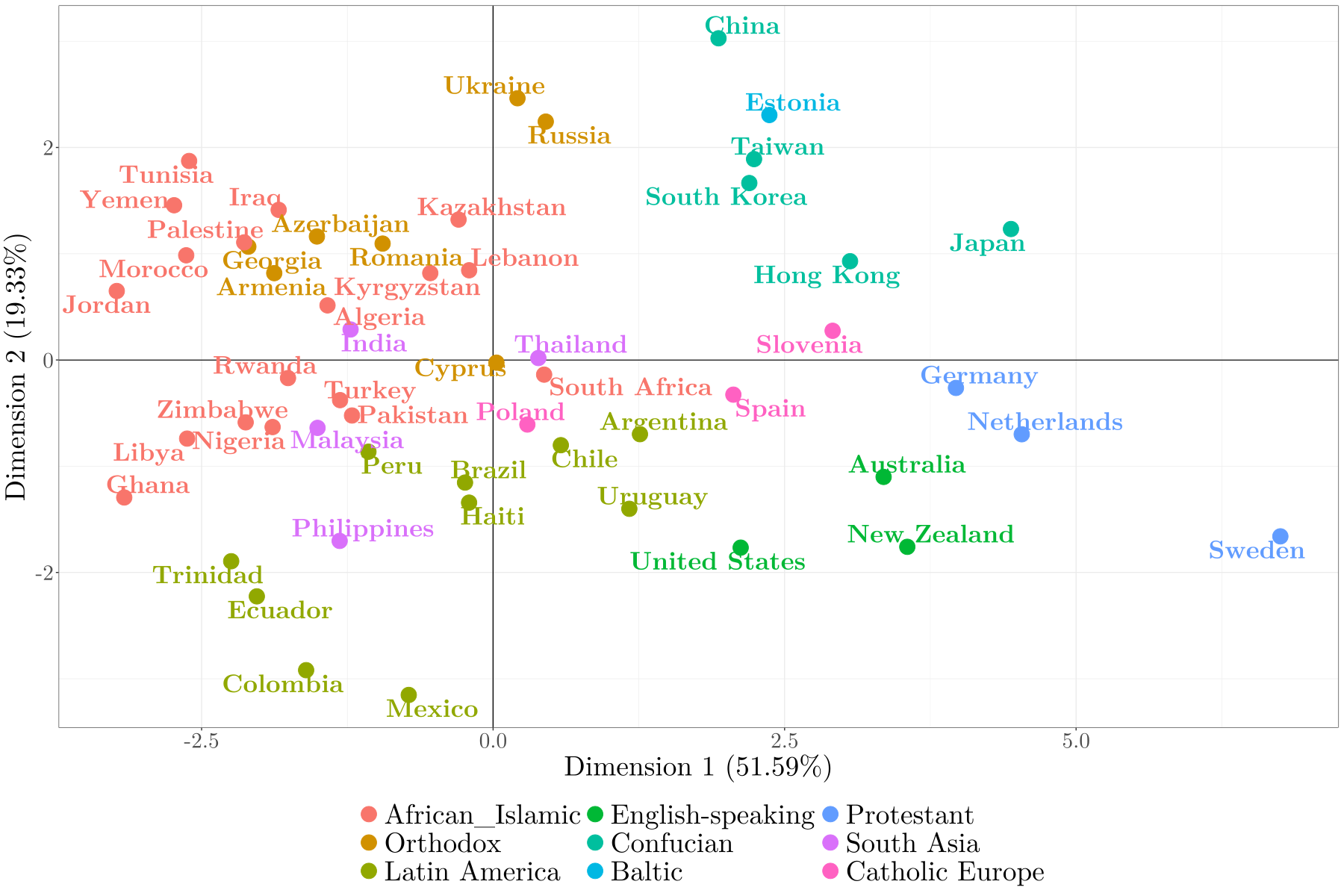}
                \label{fig:map6}
                \end{center}
                     \vspace{-0.5cm}
                {\scriptsize {\bf Note}: {\linespread{0.5}\selectfont Authors' elaborations on WVS Wave 6. Colours correspond to \cite{IngWel2005} groups. The horizontal axis corresponds to the first principal component of a PCA on the 10 variables associated to the cultural traits considered in the analysis. The vertical axis corresponds to the second principal component. Visualisation differences with respect to the original analysis are due to the rotation of axes.\par}}
            \end{figure}

In this Section, the original \cite{IngWel2005} Cultural Map is \textcolor{black}{reproduced} using the same set of questions and the same pool of countries
.
\newline
\newline
The procedure used by \cite{inglehart1997modernization} is followed. The individuals' responses collected through the questionnaire are synthesised, disregarding the higher moments 
of the distribution, in a country mean score for every country and question. These scores constitute the variables measuring the selected cultural traits.  
A \textcolor{black}{Principal Component Analysis (PCA)} is applied on the resulting 54$\times$10 countries-variables matrix. 
The first two principal components explain 71\% of the variability in the data. The interpretation of these two 
dimensions is related to variables on greater incidence. In particular, correlations between variables and 
\textcolor{black}{principal components} are captured by the coordinates in the correlation circle, displayed visually in Figure \ref{fig:pca}.
\newline
\newline
The investigation of which of the 10 cultural traits is mostly contributing to each component, that is the two dimensions (Dimension 1 and Dimension 2) defining the axes of the correlation circle, can be made in accordance with the concepts proposed by \cite{inglehart2000modernization} and \cite{IngWel2005} in the context of their Cultural Map:
\begin{itemize}
    \item Dimension 1 - First \textcolor{black}{principal component}. 
    It goes from \emph{Survival values} (on the left of the horizontal axis) to \emph{Self-expression values} (on the right of the horizontal axis), where, in short, \emph{Survival values} refer to \texttt{importance of God} (V152), low level of \texttt{trust} in people (V24) and low propensity to sign petitions (\texttt{voice} - V85); while \textit{Self-expression values} refer to greater level of tolerance about \texttt{homosexuality} (V203) and \texttt{abortion} (V204), \texttt{independence} and autonomy (Y003), low level of \texttt{national pride} (V211) and, with secondary importance, low propensity for \texttt{respect for authority} (V69).
    
    According to \cite{IngWel2005}, \emph{Survival values} give emphasis to economic and physical security and are associated to low levels of trust and tolerance, while \emph{Self-expression values} give high priority to subjective well-being, individualism, quality of life related to environmental sustainability. A rightward movement from survival to self-expression can therefore be interpreted as the transition from industrial society to post-industrial society, as well as the embracing of democratic values.
            
    \item Dimension 2 - Second \textcolor{black}{principal component}. 
    It is  mainly related to the influence of two variables: overall \texttt{happiness} (V10) and the \texttt{post-materialism} index (Y002), that, as shown in Table \ref{tab:descriptives}, are both coded in an ascending order. The top of the vertical axis corresponds to low level of happiness and  materialist societies, while the bottom corresponds to high level of happiness and post-materialist societies. Other variables have a negligible effect on the second \textcolor{black}{principal component}.
    
    In the original \cite{IngWel2005} Cultural Map, this second dimension captures the spectrum of countries' characteristics from \emph{traditional} to \emph{secular-rational values}. ``Traditional values emphasize the importance of religion, parent-child ties, deference to authority, absolute standards and traditional family values. People who embrace these values also reject divorce, abortion, euthanasia and suicide. Societies that embrace these values have high levels of national pride and a nationalistic outlook.'' A downward shift represents a movement away from traditional values and toward more secular-rational values, strongly mediated by happiness. \textcolor{black}{In our case, the \emph{traditional} vs \emph{secular values of societies} dichotomy is mediated strongly by happiness, creating a discrepancy with the \cite{IngWel2005} analysis. This may be attributed to specific choices made by \cite{IngWel2005} and not replicated in our analysis, e.g. a possible re-balancing of the data using weights provided by the WVS or a synchronization of the analyses along the longitudinal dimension of the previous WVS waves, which are considered by \cite{IngWel2005} and not by the present study}.
\newline
\end{itemize}
Using the above interpretation of the PCA components, the position of the countries in the factor\textcolor{black}{ial} map depicted in Figure \ref{fig:map6} describes a cultural topology: the further away from the origin a country is, the more the national culture of that country is explained by the interplay of the two components; the closer a country is to one particular component, the more it is explained by that particular component only. For consistency with the original \cite{IngWel2005} analysis, the hybrid geographical/religious classification of countries is maintained and displayed with different colours.
\newline
\newline
For ease of discussion, let us focus on the four quadrants of Figure \ref{fig:map6} separately. Quadrant 1, to the top-left, is the \emph{Survival-Materialist} quadrant. It includes a subset of African/Islamic countries and a subset of Orthodox countries, plus India, that lies close to the left of the axis of the first component (e.g. high \texttt{importance of God}, low \texttt{trust}, little \texttt{voice}). Tunisia and Iraq show the highest average value of \texttt{M} and \texttt{H}, respectively (with an average \texttt{M} equal to 1.73 for Tunisia and an average \texttt{H} equal to 2.255 for Iraq, in Table \ref{tab:descriptives}), while the average value of \texttt{G} is equal to 9.865 for Yemen, and that of \texttt{V} is equal to 2.857 for Azerbaijan, the highest and lowest, respectively, in Table \ref{tab:descriptives}. Quadrant 2, to the top-right, is the \emph{Self-Expression-Materialist} quadrant. It includes all Confucian countries and Estonia, Ukraine and Russia, and Slovenia. This latter country lies close to the right side of the axis of the first component (e.g. tolerance for \texttt{homosexuality} and \texttt{abortion}, low \texttt{respect for authority}, high relevance of \texttt{independence}). Ukraine and Russia, instead, stand out at the top of the second component axis (e.g. extreme \texttt{materialism} and low level of \texttt{happiness}). Quadrant 3, to the bottom-left, is the \emph{Survival-Post-materialist} quadrant. It includes the remaining African/Islamic countries (but South Africa), a subset of South Asian countries, and a subset of Latin American countries with a high level of \texttt{happiness} (indeed Mexico has the lowest average value of \texttt{H}, equal to 1.387) and, on the one hand, counterbalancing tendencies to preserve traditional values and, on the other, to be attracted by \texttt{post-materialist} values. Quadrant 4, to the bottom-right, is the \emph{Self-Expression-Post-materialist} quadrant. It includes European countries, both catholic and protestant, the disjoint of the previous subset of Latin American countries and Australia, New Zealand and the US. Sweden is at the edges of the quadrant (e.g. tolerance for \texttt{homosexuality} (average \texttt{O} equal to 8.47) and \texttt{abortion} (average \texttt{A} equal to 7.997), low relevance of the \texttt{respect for authority} as an individual value, high relevance of \texttt{independence}, \texttt{post-materialism} and a fair level of \texttt{happiness}). Finally, Cyprus is the pivot country, being positioned right a the intercept of the two orthogonal axes. Country groups tend to cluster, with some degree of overlap and some spreading over different quadrants. 

\section{Accounting for cultural traits interdependence}
\label{methods}

\textcolor{black}{As it is well summarised in \cite{taras2005cross}, the large majority of cross-cultural studies just focuses on group (national) means of cultural scores, in the tradition of \cite{hofstede1984culture} or \cite{schwartz1992universals}. In this way, higher moments of the distribution of within group scores are neglected in the measurement of international cultural distances (see also \cite{taras2009half} and \cite{kostova2021integrating} for a critique of this practice in the context of cultural distance studies).} The analysis of \cite{IngWel2005}, and the \texttt{IW index} that resulted from it, \textcolor{black}{also belongs to this tradition, being based on the comparison between country-means and the follow-up principal component analysis of these scores.
The alternative approach, proposed in this paper and discussed in the next section, 
augments existing measures of cultural distance by taking into account both the full distribution of the individual cultural traits and the interdependence between them}. This allows to capture countries that are both similar in their cultural traits and in the way these cultural traits are inter-connected with each other. For example, some of those traits could be connected in a positive or negative way because that country symbiotically expresses those values; some others may not be connected to each other because in that national culture those values are implicitly not related; finally some cultural traits may not be significantly part of the national culture for a certain country. Using a network analogy, if cultural traits are nodes and cultural traits interdependence defines the set of edges linking the different nodes, national culture is the joint set of nodes and edges, of cultural traits and their connections. If this is the case, in every specific national culture some cultural traits can appear more central while some nodes can be detached from the rest of the national cultural network. Therefore, national cultures can be similar not only because they share the same cultural traits but also because those very traits are more or less inter-connected in a similar way and form a similarly multifaceted relational structure. A similar network structure of cultural traits make countries similar in national cultures. Dissimilarity makes them structurally distant.
\newline
\newline
Since these cultural trait dependencies are not observed, statistical methods are needed to infer the underlying network of cultural values for each country. Using the previously used analogy as a concrete stepping stone in the proposed research framework, a network, \textcolor{black}{in the cases presented, is going to be composed of} 10 cultural traits, as nodes, and the discovery of edges is performed individually for each country. The next Section describes the specific method selected for network inference and highlights the innovative aspect with respect to existing approaches. In the following Section, a measure of distance between the inferred cultural networks is going to be developed.

\subsection{Inference of a country cultural network}
    \label{sec:gmodel}
In statistics, dependency between two variables can be captured by their correlation. When more than two variables are present, graphical models are a popular tool for recovering their network of dependencies \citep{lauritzen1996graphical}. In particular, Gaussian graphical models (Ggm) measure dependencies between two variables in terms of their partial correlation, i.e. the partial correlation of the two nodes conditional on 
the remaining ones. Compared to the simpler approach of measuring pairwise correlations, partial correlation allows to filter the more direct dependencies from the indirect ones, i.e those that are mediated by other nodes. Since the responses to the 10 selected questions of the WVS Wave 6 are ordinal, and therefore not normally distributed, Gaussian copula graphical models (Gcgm) is considered in the analysis. 
\newline
\newline
Going into the details of a Gcgm, let $\bm{Y} = (Y_{1},\ldots, Y_{p})$ be the vector of cultural traits  for a certain country, measured by the responses to $p$ survey questions (for simplicity, the notation for the country is omitted in the description of the model, as inference is conducted individually for each country). 
\textcolor{black}{Formally, in a Gcgm \citep{song00}, the joint distribution of $\bm{Y}$ is modelled by
\begin{equation}
        \label{eq:model}
        P(Y_{1} \leq y_{1},\ldots, Y_{p} \leq y_{p}) 
        = \Phi_{p}(\Phi^{-1}(F_{1}(y_1)), \ldots, \Phi^{-1}(F_{p}(y_p))|\bm{R}),
    \end{equation}
     where $F_{j}$ is the marginal cumulative distribution function of the $Y_{j}$ cultural trait, with $j=1,\ldots,p$, $\Phi(\cdot)$ is the univariate cumulative distribution function of a standard Gaussian distribution and $\Phi_{p}$ is the cumulative distribution function of the $p$-dimensional multivariate normal with mean zero and correlation matrix $\bm{R}$. In the context of copulas, this is the case of setting the copula function to the mapping $C: [0,1]^p \rightarrow [0,1]$ defined by
    \[C(u_1,\ldots,u_p) 
        = \Phi_{p}(\Phi^{-1}(u_1), \ldots, \Phi^{-1}(u_p)|\bm{R}),
    \]
    which is indeed a copula, as it is a continuous distribution with uniform marginals, and  complementing it with the $Y_j$ marginal distributions to describe the joint distribution of $\bm{Y}$, as
    \[P(Y_{1} \leq y_{1},\ldots, Y_{p} \leq y_{p}) = C(F_{1}(y_{1}),\ldots,F_{p}(y_{p})~| \bm{R}).\]
    The procedure 
    for the construction of the joint distribution of $\bm Y$ with given marginals is guaranteed by Sklar's theorem \citep{sklar59}, which is central to the theory of copulas. From the cumulative distribution function in (\ref{eq:model}), the joint probability distribution can be derived. Since the $Y_j$ variables are discrete, the joint probability mass function is given by 
\begin{equation}
\label{eq:pmf}
p(y_1,\ldots, y_p) = \sum_{j_1=1}^2\cdots\sum_{j_p=1}^2(-1)^{\sum_{h=1}^p j_{h}}\Phi_{p}(\Phi^{-1}(u_{1j_{1}}), \ldots, \Phi^{-1}(u_{pj_{p}})|\bm{R}),
\end{equation}
where $u_{j1} = F_j (y_j)$ and $u_{j2} = F_j ({y_j}^{-})$, with $F_j ({y_j}^{-})$ the left-hand limit of $F_j$ at $y_j$ \citep{song00}.
 }
\newline
\newline
When it comes to modelling dependencies in the data, it is interesting to note how a Gcgm  couples, as indeed the denomination \emph{copula} suggests, the information from the individual cultural traits' distributions 
with the network component of their interdependencies. The latter is captured in the latent Gaussian space of $\bm Z = (Z_{1}=\Phi^{-1}(F_{1}(Y_1))),\ldots, Z_{p}=\Phi^{-1}(F_{p}(Y_P)))$ by the inverse of the correlation matrix $\bm{K} = \bm{R}^{-1}$. This matrix,  typically called the precision or concentration matrix, has a special role in a Ggm, as it is uniquely associated to the underlying network, which is here interpreted as a conditional independence graph. In particular, denoting with $k_{ij}$ the ($i$,$j$) element of the $\bm{K}$ matrix:
\begin{equation*}
Z_{i} \indep Z_{j} ~|~ \bm{Z}_{-ij} \quad \mbox{if and only if} \quad k_{ij} = 0, \quad i \neq j,
\end{equation*}
    that is $Z_i$ is conditionally independent of $Z_j$ given all the other variables ($\bm{Z}_{-ij}$) (i.e. a missing link in the graph) if and only if the corresponding $(i,j)$ element of the precision matrix is zero \citep{lauritzen1996graphical}.  Equivalently, conditional independence and a zero precision value can be shown to correspond to a zero partial correlation, as
    \begin{equation}
            \gamma_{ij} = \rho(Z_{i},Z_{j} ~|~ \bm{Z}_{-ij}) = -\frac{k_{ij}}{\sqrt{k_{ii} k_{jj}}}, \quad i \neq j.
    \label{eq:corr}
    \end{equation}
In 
comparison with traditional methods, it is of interest to note the difference with  the \cite{IngWel2005} approach for measuring cultural distances. Firstly, the \cite{IngWel2005} approach does not consider the dependency component of cultural traits, essentially assuming a diagonal correlation matrix $\bm{R}$. Secondly, the \cite{IngWel2005} approach works with aggregate measures of the data, by taking the mean of each cultural trait, thus essentially considering the expected value associated to each marginal distribution $F_j$ rather than the distribution in its entirety. Evidence in Table \ref{tab:descriptives} is exemplificative of the potential bias of this procedure. 
\newline
\newline
Inference in a Gcgm is traditionally made of two tasks: estimation of the marginal distributions, one for each cultural trait, and estimation of the underlying multivariate distribution of their inter-dependencies. The first task is relatively simple and is typically tackled with the use of empirical distributions as estimates of the true distributions. In order to avoid problems with zero probabilities, in this paper, the Bayesian estimates of the bin frequencies via the Dirichlet-multinomial pseudocount model is used, assuming a uniform prior distribution across the categories for each cultural trait.  The second task is instead challenging, particularly in high dimensions, and can be itself split into two sub-tasks: \emph{parameter estimation}, that is estimation of the precision matrix $\bm{K}$, and \emph{model selection}, that is selection of a graph where some edges may be missing. In a frequentist framework, these two tasks are computed separately: given a graph, the precision matrix is estimated by constrained maximum likelihood, whereas model selection criteria based on the model likelihood and model complexity (in this case number of edges) are subsequently used to select a single optimal graph \citep{lauritzen1996graphical}. In contrast to this, a Bayesian approach, such as the one considered in the present case, allows to account simultaneously for uncertainty both at the level of graph inference and precision matrix estimation, by returning their full posterior distribution.
\newline
\newline
\textcolor{black}{As with Bayesian procedures, priors need to be defined on the parameters of interest, in this case the precision matrix $\bm{K}$ and the graph $G$. These are then combined with the likelihood function to produce posterior distributions. Starting with the likelihood function, this is given by Equation (\ref{eq:pmf}) evaluated on the observed data $\bm{Y}$.  For inference purposes, it is convenient  to write the likelihood in terms of the latent data $\bm{Z}$. Following \cite{mohammadi2017bayesian} and, more broadly, the extended rank likelihood approach of \cite{hoff2007extending}, the key observation is that, given  $\bm{Y}$ and the fact that the marginal distributions $F_j$ are non-decreasing, $\bm{Z}$ is constrained to take values in certain intervals. In particular, 
these are defined by
\begin{equation*}
        \mathcal{D}(\bm{Y}) = \{ \bm{Z} \in \mathbb{R}^{n \times p} : \Phi^{-1}(F_j ({y_{ij}}^{-})) < z_{ij} < \Phi^{-1}(F_j ({y_{ij}})),  i = 1,\ldots,n; j = 1,\ldots,p\},
    \label{eq:LeU}
    \end{equation*}
    where $n$ is the number of observations.  It is interesting to note that, for missing data, which is rather common for survey data (e.g., see Table \ref{tab:descriptives}), the corresponding interval is simply set to  $(-\infty, \infty)$. Since the event $\bm{Z} \in \mathcal{D}(\bm{Y})$ occurs whenever $\bm{Y}$ is observed, the likelihood can be written as
    \begin{eqnarray}
P(\bm{Y} ~|~ \bm{K}, G, F_1,\ldots,F_p) &=& P(\bm{Y},\bm{Z} \in \mathcal{D}(\bm{Y}) ~|~ \bm{K}, G, F_1,\ldots,F_p) \\ \nonumber
&=&  P(\bm{Z} \in \mathcal{D}(\bm{Y}) ~|~ \bm{K}, G, F_1,\ldots,F_p)P(\bm{Y}~|~ \bm{Z} \in \mathcal{D}(\bm{Y}), \bm{K}, G, F_1,\ldots,F_p).
    \label{eq:likyz}
    \end{eqnarray} 
Given some fixed marginal distributions, e.g., typically estimated offline using the empirical distributions \citep{mohammadi2017bayesian} or, like in the present case, using Bayesian estimates,  the only part of the observed data likelihood that depends on $\bm{K}$ and $G$ is given by:
    \begin{equation}
        P(\bm{Z} \in \mathcal{D}(\bm{Y}) ~|~ \bm{K}, G) = \int_{\mathcal{D}(\bm{Y})} P(\bm{Z} | \bm{K}, G)~ d\bm{Z}
    \label{eq:hoff}
    \end{equation} 
    where $P(\bm{Z} | \bm{K}, G)$ is the profile likelihood in the Gaussian latent space, that is
    \begin{equation*}
         P(\bm{Z} | \bm{K}, G) \propto |K|^{n/2} \exp \biggl\{-\frac{1}{2} \mbox{Trace}(\bm{KU}) \biggr\}
    \end{equation*}
    with $\bm{U} = \bm{Z}^{T} \bm{Z}$ denoting the sample moment. This is a useful representation of the model when it comes to Bayesian inference, as it essentially resorts to sampling the latent data $\bm{Z}$ from a multivariate normal distribution, truncated on the intervals $\mathcal{D}(\bm{Y})$ \citep{mohammadi2017bayesian}. 
The likelihood function is finally combined to prior distributions to give the posterior distribution:
    \begin{equation*}
        P(\bm{K}, G ~|~ \bm{Z} \in \mathcal{D}(\bm{Y})) \propto P(\bm{Z} \in \mathcal{D}(\bm{Y}) ~|~ \bm{K}, G)P(\bm{K}|G)P(G). 
    \end{equation*}
Following \cite{mohammadi2017bayesian}, an Erd\"{o}s-R\'{e}nyi random graph with a noninformative 0.5 prior probability on each link is considered for $P(G)$, while,
given a graph $G$, a  G-Wishart distribution is considered for the precision matrix $\bm{K}$. This is given by:
\begin{equation*}
         P(\bm{K} | G) = \frac{1}{I_{G}(b,\bm{D})} |\bm{K}|^{(b-2)/2} \exp \biggl\{-\frac{1}{2} \mbox{Trace}(\bm{D}\bm{K}) \biggr\},
    \end{equation*}
where $I_{G}(b,\bm{D})$ is a normalizing constant. For the real data analysis, $b$ is set to 3 and the symmetric positive definite matrix $\bm{D}$ is taken as the identity matrix $\mathbb{I}_p$.
}
\newline
\newline
While computational approaches to sample from the posterior distribution in the context of graphical models were prohibitive until not long ago, recent fast implementations have been proposed based on Birth-Death Markov Chain Monte Carlo (BDMCMC) methods both for the case of a Ggm, for Gaussian data \citep{mohammadi2015bayesian}, and a Gcgm, for discrete data \citep{dobra2011copula,mohammadi2017bayesian, mohammadi2019bdgraph}. 
After convergence of the MCMC routine, the posterior on the graph space is returned, whereby each graph is given a weight, corresponding to the time the process visited that graph. Rather than selecting a single optimal graph, as in the frequentist approach, the posterior distribution can be used to calculate  estimates of quantities of interest by Bayesian averaging. In particular, for the present analysis,  the \emph{Posterior Edge Inclusion Probabilities} are considered. These are given by:
\begin{equation}
\pi_e=P(e \in E~|\bm{Y}) = \frac{\sum_{t=1}^{N} 1(e \in G^{(t)}) W(\bm{K}^{(t)})}{\sum_{t=1}^{N} W(\bm{K}^{(t)})},
\label{eq:postprob}
\end{equation}
where $E$ is the set of edges, $N$ is the number of MCMC iterations and $W(\bm{K}^{(t)})$ is the waiting time for the graph $G^{(t)}$ with the precision matrix $\bm{K}^{(t)}$.    
Similarly, one can obtain a single precision matrix $\bm{K}$  by taking an average of all the precision matrices sampled during MCMC, again weighted by the graph probabilities. From this, the \emph{Partial Correlation matrix} can be derived using Equation (\ref{eq:corr})
. This matrix is a more informative output than the posterior edge probability as it contains both the intensity and the sign of the relationships between cultural traits.
\newline
\newline
While these are two direct outputs from the inferential scheme, it is also interesting to note how the posterior distribution of any other quantity of interest can also be obtained by exploiting the full sequence of graphs sampled from the posterior distribution, without the need for additional re-sampling procedures such as bootstrap. This is also put to good use in the subsequent analysis, where node and graph centrality measures across the different countries are going to be calculated and analysed.  

\subsection{A unifying index of cultural distance}
\label{sec:distancelist}
With the method just described, a network is inferred for each country. Each model provides a measure of the inter-connectedness within each country, via the precision matrix and the graph, and of the distributions of the individual cultural traits, via the marginal distributions.  In this Section, these different components of the model are integrated into the calculation of one measure of cultural distance between countries. The resulting measure should account both for the distance between the inferred networks, capturing how countries' cultural traits are inter-connected, and for the distance between the marginal distributions of individual traits, representing country's attitudes to the individual cultural traits. 
\newline
\newline
\textcolor{black}{A measure of distance between the models associated to two countries can be derived by considering the distance between their corresponding joint probability distributions (Equation \ref{eq:pmf}), evaluated using the Bayesian estimates of $\mathbf{K}$ and of the marginal distributions for each country.  A common measure of divergence between two distributions is the Kullback-Leibler ($KL$) divergence \citep{kullback1951information}. In the specific case of the Gcgm, denoting with $\mathbf{F}^{(c)}=(F_1^{(c)},\ldots,F_p^{(c)})$ the cumulative distribution functions of the cultural traits for a country $c$, and with $\mathbf{K}^{(c)}$ the estimated precision matrix of their dependencies, the KL divergence between the Gcgm distributions for country $m$ and $l$ is given by
\[KL(m ~||~ l)  = \sum_{y_1,\ldots, y_p} p(y_1,\ldots, y_p | \mathbf{K}^{(m)},\mathbf{F}^{(m)} ) \log\Big(\frac{p(y_1,\ldots y_p | \mathbf{K}^{(m)},\mathbf{F}^{(m)})}{p(y_1,\ldots, y_p | \mathbf{K}^{(l)},\mathbf{F}^{(l)})}\Big),\]
with the joint distribution $p( ~\cdot ~|~ \mathbf{K},\mathbf{F})$ evaluated
at all distinct instances $(y_1,\ldots, y_p)$ of the multivariate vector $(Y_1,\ldots,Y_p)$. Calculating this measure for any pair of countries is time consuming in high dimensions (i.e., the case of a large $p$ or a large number of categories per cultural trait). Moreover, from a practical point of view, it would be useful to distinguish the contribution to the divergence coming from the marginal distributions and that from the network components. To this aim, an approximate divergence is considered, which is exact only in the case of continuous marginals. Indeed, in this case  the $KL$ divergence decomposes into the $KL$ divergence between the marginal densities and the $KL$ divergence between the copula Gaussian densities \citep{lasmar14}.}
\newline
\newline
\textcolor{black}{In order to derive this measure and appreciate the level of approximation, note that the intervals constructed as in $\mathcal{D}(\bm{Y})$ but evaluated on all possible instances $(y_1,\ldots, y_p)$ of the multivariate vector $(Y_1,\ldots,Y_p)$, partition the $p$-dimensional $\bm{Z}$ space into $T_1\times T_2\times \dots \times T_p$ rectangular cubes, where $T_j$ is the total number of categories for the $j$-th survey question $Y_j$. Approximating the integral in (\ref{eq:hoff}) via the multivariate mean value theorem \citep{abegaz15} and assuming a large number of categories, one obtains
\[
P(Y_1=y_1,\ldots,Y_p=y_p ~|~ \bm{K}, G, F_1,\ldots,F_p) \approx  c(F_1(y_1),\ldots,F_p(y_p))\prod_{i=1}^pf_j(y_j)\Delta V(y_1,\ldots,y_p),
\]
with $f_j$ the marginal probability distributions of the $j$-th cultural trait, $c$ the copula density
\[c(u_1,\ldots,u_p)=|\bm{R}|^{-1/2}\exp\Big\{-\frac{1}{2}\mathbf{q}^t(\mathbb{I}_p-\bm{R}^{-1})\mathbf{q}\Big\},\]
with $q_j=\Phi^{-1}(u_j)$, and  $\Delta V(y_1,\ldots,y_p)$ the volume of the p-dimensional cube generated by observation $(y_1,\ldots, y_p)$.
Denoting with  $\mathbf{f}^{(k)}=(f_1^{(k)},\ldots,f_p^{(k)})$ the marginal probability distributions of the $p$ cultural traits in country $k$, and using now a similar derivation to that adopted for continuous marginals \citep{lasmar14},  the $KL$ divergence between two countries $m$ and $l$ is finally approximated by
\begin{eqnarray*}
KL(m ~||~ l)&\approx&\sum_{i=1}^p KL(f_i^{(m)} || f_i^{(l)}) + KL(c(\mathbf{F}^{m}; \mathbf{K}^{(m)}) || c(\mathbf{F}^{l};\mathbf{K}^{(l)} )) = \\
&=&\sum_{i=1}^p \sum_{k=1}^{T_i} f^{(m)}_{i}(t_k) \log \biggl(\frac{f^{(m)}_{i}(t_k)}{f^{(l)}_{i}(t_k)}\biggl) + 0.5 \Big( \mbox{Trace}(\mathbf{K}^{(l)}(\mathbf{K}^{(m)})^{-1})+\log \frac{|\mathbf{K}^{(m)}|}{|\mathbf{K}^{(l)}|}-p\Big),
\end{eqnarray*}
 where $f^{(l)}_{i}(t_k)$ is the probability associated to the category $t_k$ for country $l$.   As well as the anticipated decomposition of this measure into the $KL$ divergence between the marginal probability distributions and the $KL$ divergence between the network components, it is interesting to note how the $KL$ divergence between the marginal distributions further decomposes into the $KL$ divergence between the distributions of the individual cultural traits. Note also how measuring distances at the level of distributions, rather than means, provides a natural scaling with respect to the specific coding used for each variable (the $t_k$ values) and the number of categories associated to each question ($T_i$). Moreover, as mentioned earlier, the use of Bayesian estimates of the bin frequencies avoids the problem with zero probabilities in the formula above.}
\newline
\newline
As the $KL$ divergence is not symmetric, the sum of the two (approximate) divergences associated to two countries, also called the Jeffreys' divergence, is considered as the final index of distance between country $m$ and $l$:
\begin{eqnarray}
JD(m,l)\!&\!=\!& KL(m || l) + KL(l || m) = \\ \nonumber
&\!=\!& \sum_{i=1}^p \sum_{k=1}^{T_i} (f^{(m)}_{i}(t_k)-f^{(l)}_{i}(t_k)) \log \biggl(\frac{f^{(m)}_{i}(t_k)}{f^{(l)}_{i}(t_k)}\biggl) + \dfrac{\mbox{Trace}(\mathbf{K}^{(l)}(\mathbf{K}^{(m)})^{-1})+\mbox{Trace}(\mathbf{K}^{(m)}(\mathbf{K}^{(l)})^{-1})}{2}-p.
\label{eq:jeffrey}
\end{eqnarray}
This is operationalized in the calculation of the \texttt{JD index} of cultural distance in Section \ref{sec:distresults}.

\section{Measuring cultural distance from the WVS}
\label{sec:results}

\subsection{The inferred networks of cultural values}
\label{sec:netresults}
\begin{figure}[!ht]
        \begin{center}
        \caption{\small{\textit{Sinaplots of the topological properties of country-networks.}}}
            \includegraphics[width=\textwidth]{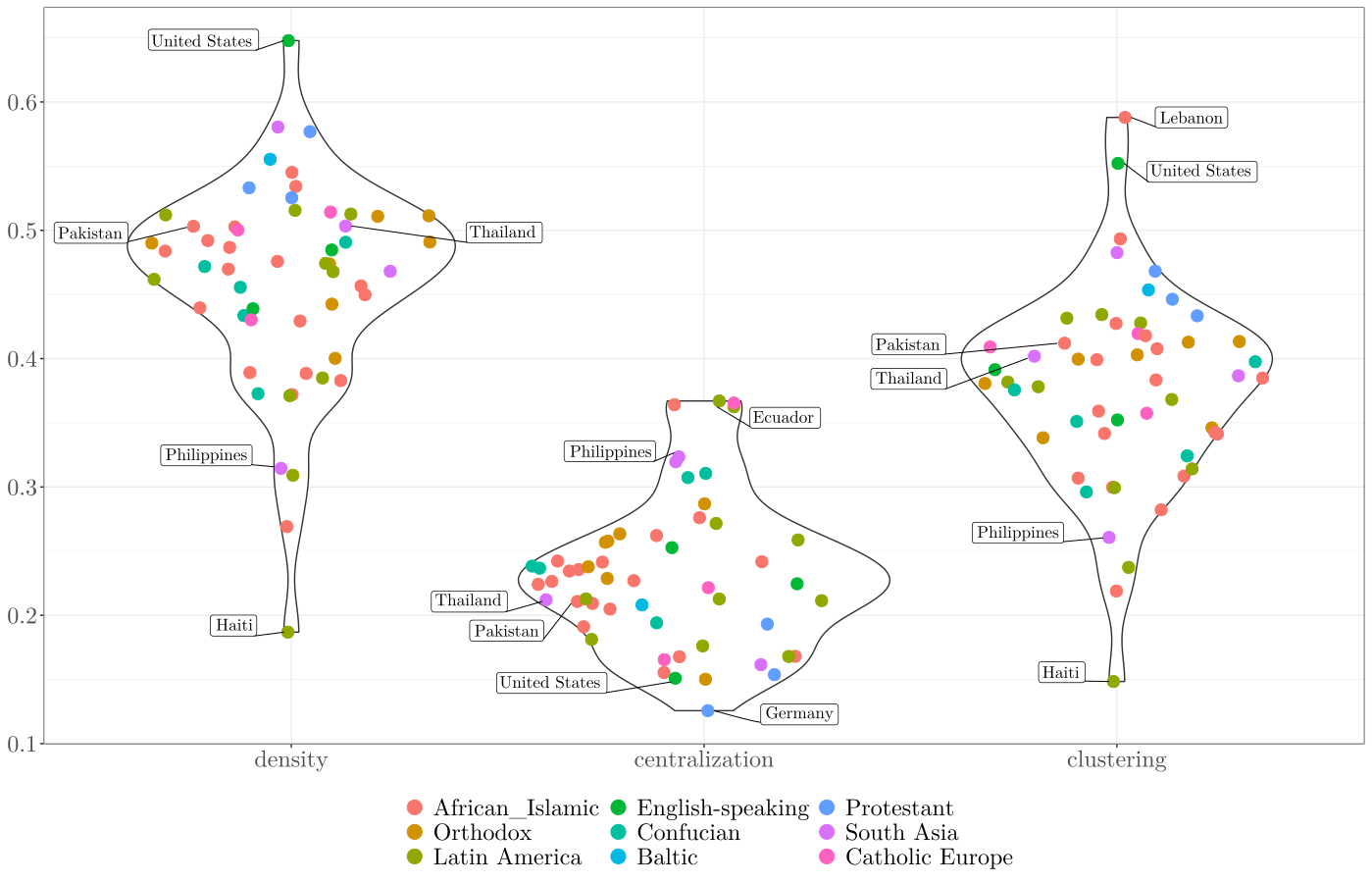}
        \label{fig:netsummaries}
        \end{center}
             \vspace{-0.5cm}
        {\scriptsize {\bf Note}: {\linespread{0.5}\selectfont Authors' elaborations on WVS Wave 6. Colours correspond to \cite{IngWel2005} groups. The Figure summarises the 54 cultural networks in terms of overall density (left), network centralization (middle) and clustering coefficient (right). Each measure is calculated with edge weights given by the edge posterior probabilities. \textcolor{black}{The sinaplots} 
        show the empirical probability distribution of the network statistics across the 54 countries, smoothed by an optimal kernel. \par}}
\end{figure}

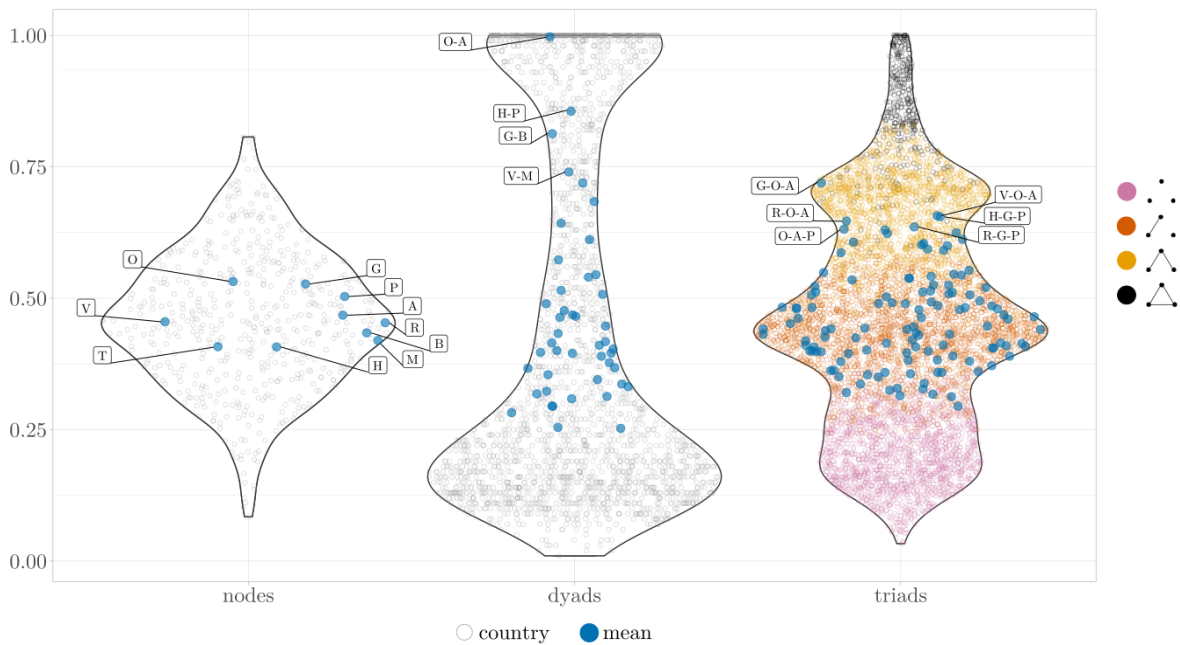
\begin{figure}[!hb]
        \begin{center}
        \caption{\small{\textit{Sinaplots of the nodes centrality, edges probabilities and triads}}}
            \includegraphics[width=\textwidth]{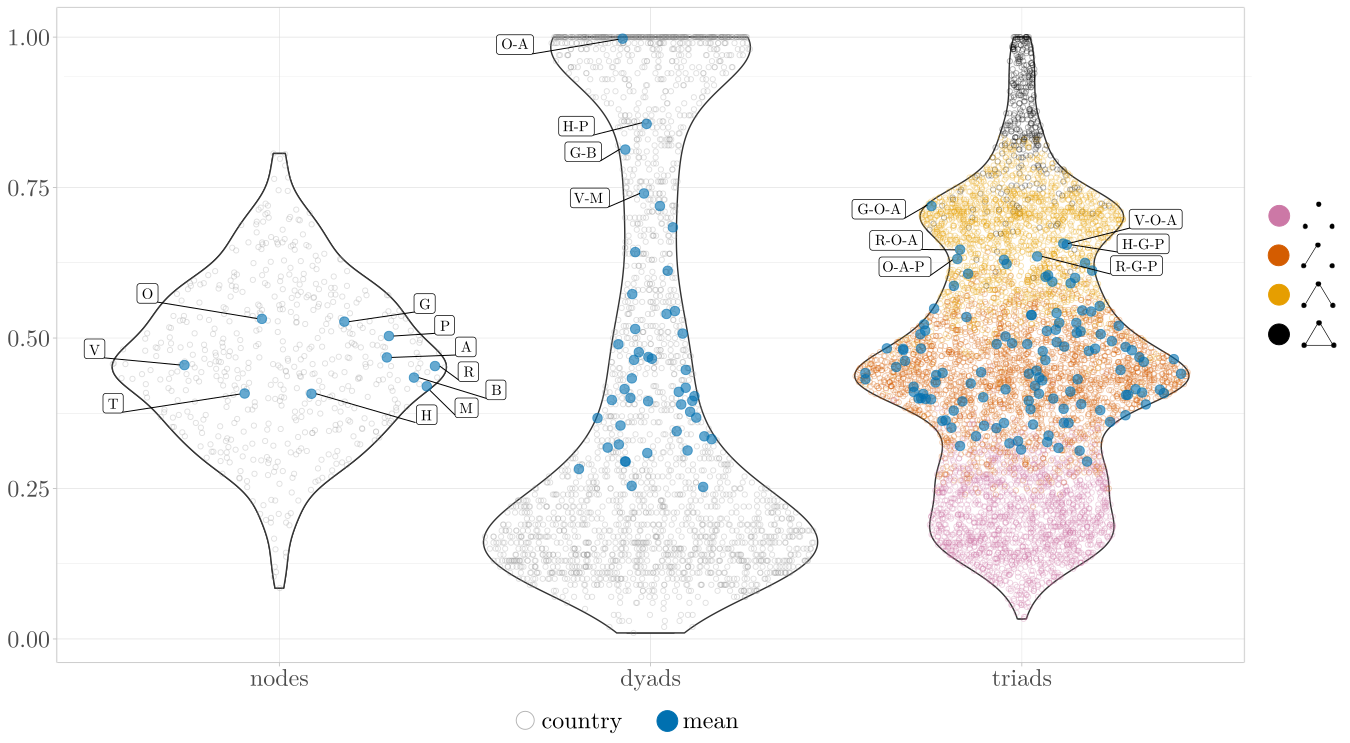}
        \label{fig:nodedyadtriad}
        \end{center}
             \vspace{-0.5cm}
                {\scriptsize {\bf Note}: {\linespread{0.5}\selectfont Authors' elaborations on WVS Wave 6. The Figure summarises the characteristics of the 10 cultural traits in the countries' probability cultural networks. The node level is described in term of weighted degree centrality (left); the dyad level by the value of the edge probabilities (middle); and the triadic configuration level by the overall weight of each possible triad, where the latter is given by the average value of the corresponding edge probabilities (right). Country values are depicted by grey circles, and average level across the 54 countries are depicted by filled blue dots. Triads are coloured according to four different configurations: rose circles indicate an open triangle (\hspace{-0.1cm}
\begin{tikzpicture}[scale=0.60]
\filldraw [black] (0,0) circle (2pt)
(0.25,0.25) circle (2pt)
(0.5,0) circle (2pt);
\end{tikzpicture} \hspace{-0.1cm}), 
orange circles indicate a triad with just one relevant side (\hspace{-0.1cm}
\begin{tikzpicture}[scale=0.60]
\filldraw [black] (0,0) circle (2pt)
(0.25,0.25) circle (2pt)
(0.5,0) circle (2pt);
\draw (0,0) -- (0.25,0.25);
\end{tikzpicture} \hspace{-0.1cm}),
yellow circles indicate a triad with two relevant sides (\hspace{-0.1cm}
\begin{tikzpicture}[scale=0.60]
\filldraw [black] (0,0) circle (2pt)
(0.25,0.25) circle (2pt)
(0.5,0) circle (2pt);
\draw (0,0) -- (0.25,0.25);
\draw (0.25,0.25) -- (0.5,0);
\end{tikzpicture} \hspace{-0.1cm}),
and black circles indicate a closed triangle (\hspace{-0.1cm}
\begin{tikzpicture}[scale=0.60]
\filldraw [black] (0,0) circle (2pt)
(0.25,0.25) circle (2pt)
(0.5,0) circle (2pt);
\draw (0,0) -- (0.25,0.25);
\draw (0.25,0.25) -- (0.5,0);
\draw (0.5,0) -- (0,0);
\end{tikzpicture} \hspace{-0.1cm}). The colour associated to each triad, depends on the level of the posterior edge probability of every link 
above the 0.5 cutoff.\par}}
    \end{figure}

The approach described in the previous Section is now applied to the WVS answers for each of the  54 countries in order to infer the dependencies among the $p$=10 cultural traits and to give evidence to the different cultural structures emerging. A generous value of 2 million iterations for the MCMC algorithm is set, and 1 million of these is retained as burn-in. All the estimates are calculated on the remaining 1 million iterations. In particular, the Bayesian inferential procedure returns a posterior probability $\pi_e$ of each pair of nodes being connected in each country (from Equation (\ref{eq:postprob})) and an estimate of the $\bm{K}$ precision matrix associated to each country, from which a partial correlation matrix can be obtained using Equation (\ref{eq:corr}).
\newline
\newline
This Section provides an exploratory analysis of the inferred cultural networks. In order to schematise the procedure adopted, the description firstly concentrates on the edge probabilities and then it considers the sign of the edge strengths that 
comes from the partial correlations $\gamma_{ij}$. 
Figure \ref{fig:netsummaries} plots network summaries \citep{wasserman1994social} for each country, all calculated as weighted, with weights given by the edge probabilities, and all with a natural support between 0 and 1. In particular, the \textcolor{black}{left sinaplot (a combination of a violin plot and the jittered data points of a strip plot)} 
reports the densities of the country-networks, that is the average of the edge probabilities of a country's inferred network, and the resulting empirical distribution for all countries. The sinaplot shows a slight sign of bi-modality and a large heterogeneity among countries, with some, like Haiti, having a very low connection among the cultural traits and others, like the United States, being characterized by a very dense cultural network. Denser networks correspond to more complex and coherent national cultures, where many cultural traits are interdependent, and the relative relevance of a cultural trait largely depends also on the correlation with the other traits. Denser structures can also give rise to different cultural configurations, made of the  aggregation of different sub-graphs, offering a richer set of possible combinations of the factorisation of the network \citep{lauritzen1996graphical}. 
\newline
\newline
The middle sinaplot reports, for each country, the coefficient of variation of the weighted node degrees. This can be taken as a measure of centralization of a country's network, that is a measure of how much individual cultural traits may take a central position in the wired structure of a national culture. Of course, centralization is negatively correlated with density and countries such as the United States and Germany are characterised by a very low centralization;  Pakistan and Thailand show a middle level of centralization; and The Philippines and Ecuador are among the countries for which the degree distribution of the cultural traits is more asymmetric, with a limited number of cultural traits having many links with high posterior edge probabilities, and many cultural traits associated to low posterior edge probabilities. 
\newline
\newline
Finally, the more symmetric  and uni-modal right \textcolor{black}{sinaplot}, reporting the average weighted clustering coefficient \citep{onnela2005intensity} per country, shows a large heterogeneity also in the level of inter-connected triplets of cultural traits, with countries like Haiti and the United States at opposite extremes. Lebanon is the country with the highest number of cliques formed by  triangles of cultural traits in which all sides have very high posterior probabilities. Pakistan and Thailand show an average level of clustering, while the cliques in the network of The Philippines are limited, indicating a low number of possible cultural configurations.
\newline
\newline
On the other hand, Figure \ref{fig:nodedyadtriad} shows the potential differences/commonalities between the 54 country-networks, by highlighting nodes, dyads or triads (the simplest sub-graph configurations that are taken into account) that may be central or highly prevalent across the 54 networks. In particular, the left plot reports the weighted degree of each cultural trait, at the country level (grey circles) and at an average level across the 54 countries (filled blue dots). The node with the highest level of weighted degree is \texttt{O}, showing that the opinions on the justification of homosexuality are central in a large majority of national cultural networks. The highest level of centrality of \texttt{O} (0.799, at the top limit of the distribution) is for the case of Romania, while the lowest level (0.2, in the lowest part of the distribution) is for the case of Haiti. 
\newline
\newline
The same inspection can be done for all others cultural traits. Haiti, being characterised by a highly disconnected cultural network (see the density sinaplot in Figure \ref{fig:netsummaries}), is the country with the lowest level of node centrality for all cultural traits, with the exception of \texttt{M}, for which Trinidad reaches the lowest level, and \texttt{H}, for which Ecuador gets the lowest. This evidence contributes to the explanation of the position of the two countries in the Cultural Map in Figure \ref{fig:map6}, with both countries in the third (bottom-left) quadrant, far from the vertical axis of Dimension 2, which is influenced by \texttt{happiness} and \texttt{post-materialism}. As far as the highest centrality goes, \texttt{T} and \texttt{M} are highly central in the case of Sweden (0.59 and 0.66, respectively), while \texttt{P} and \texttt{G} are so in the case of the United States (0.72 and 0.81). The highest value of centrality for \texttt{H} is in the case of Estonia (see its position in Figure \ref{fig:map6}). While the highest for \texttt{R} is  Rwanda (0.81), for \texttt{V} is  Slovenia (0.74), for \texttt{A} is  South Africa (0.70), and for \texttt{B} is  Colombia (0.65).
\newline
\newline
The middle \textcolor{black}{sinaplot} shows the edge probabilities for each possible dyad, and their average across the 54 networks. Here it is evident how certain cultural traits, such as the tolerance towards \texttt{homosexuality} and \texttt{abortion}, are strongly relevant across all networks, with the average edge probability taking the maximum value of 1. The link between \texttt{O} and \texttt{A} is present in all countries, with no exceptions, as one would expect from the distributions shown in Table \ref{tab:descriptives}. Less extreme cases, but still relevant ones, are the dyad \texttt{happiness} and \texttt{national pride}, the one referring to the link between the \texttt{importance of God} and \texttt{obedience vs independence}, and the one between \texttt{voice} and \texttt{post-materialism}. The dyads \texttt{O-A}, \texttt{H-P}, \texttt{G-B}, and \texttt{V-M} are just examples of the most prevalent simple configurations shared by the majority of national cultural networks. Indeed, the blue labelled dots in the middle \textcolor{black}{sinaplot} in Figure \ref{fig:nodedyadtriad} represent the average values of these dyads across the countries. The blue dots with the lowest levels are instead \texttt{H-B} and \texttt{H-A}, but the values at the country level (grey circles) can be significantly below or above the average value, highlighting the specificity of national cultures (e.g. the \texttt{H-A} dyad in Yemen has an edge probability which is more than twice the average one, and in the case of Kyrgyzstan the edge probability is five times lower than the average one). The empirical distribution represented by the \textcolor{black}{sinaplot} is highly bi-modal due to the overwhelming relevance of  \texttt{O-A} and the very low edge probability of many dyads at the country level.
\newline
\newline
Finally, the right \textcolor{black}{sinaplot} reports the average edge probability for each potential triad of cultural traits, again at country (coloured circles) and average (filled blue dots) level. 
The empirical distribution is evidently multi-modal, with modes referring to different types of triangular configurations. Closed triangles, depicted by black circles, are triadic structures where all sides have very high edge probabilities. No average level lies in this area, showing that there is no triad that is omnipresent in all national cultural networks. Some triads are characteristic of some country cases: \texttt{H-G-P} reaches the maximum possible value in edge probabilities in the case of the United States, whereas \texttt{O-A-M} gets the maximum in the case of Russia.
The triad including the \texttt{importance of God}, justification of \texttt{homosexuality}
and justification of \texttt{abortion}, is the one with the highest average edge probability of 0.668, the other triads that include the dyad \texttt{O-A}, such as \texttt{V-O-A}, \texttt{R-O-A}, \texttt{O-A-P} score highly too, together with \texttt{H-G-P}, that includes \texttt{H-P}, and \texttt{R-G-P}. All these average triads lie in between the yellow and the orange areas. The former is characterized by open triangles, with only two high level edge probabilities out of the three, while the latter is characterized by triangles with only one of the three sides with a high edge probability. As before, there is heterogeneity across countries, with 
some configurations particularly relevant in some countries but nearly absent in others, such as \texttt{H-G-R} in Rwanda, with an associated average value of 1 (black area), but a very low probability  in the case of Cyprus, with an associated value ten times lower (pink area).
\newline
\newline
\begin{table}[ht]
\begin{center}
  \caption{\small{\textit{Exploratory analysis of the 54 national cultural networks}}}
  \begin{scriptsize}
    \begin{tabular}{m{1.75cm}  m{1cm} m{1cm} 
    m{1cm}  m{1cm}  m{1cm} 
    m{1cm}  m{1cm}  m{1cm}}
    \hline
    \textbf{dyads}\!& & & & & & & & \\[-1.5ex]
            & \multicolumn{2}{c}{\textbf{\texttt{H}---\texttt{P}}} & \multicolumn{2}{c}{\textbf{\texttt{V}---\texttt{M}}} & \multicolumn{2}{c}{\textbf{\texttt{G}---\texttt{B}}} & \multicolumn{2}{c}{\textbf{\texttt{O}---\texttt{A}}} \\
          & \multicolumn{1}{c}{
                \begin{tikzpicture}[scale=0.40]
                    \Vertex[size=0.1, color=white]{A} 
                    \Vertex[size=0.1, color=white, x=2]{B}
                    \Edge[color=red, lw=1pt](A)(B)
                \end{tikzpicture}
            } & \multicolumn{1}{c}{
                \begin{tikzpicture}[scale=0.40]
                    \Vertex[size=0.1, color=white]{A} 
                    \Vertex[size=0.1, color=white, x=2]{B}
                    \Edge[color=green, lw=1pt](A)(B)
                \end{tikzpicture}
            } & \multicolumn{1}{c}{
                \begin{tikzpicture}[scale=0.40]
                    \Vertex[size=0.1, color=white]{A} 
                    \Vertex[size=0.1, color=white, x=2]{B}
                    \Edge[color=red, lw=1pt](A)(B)
                \end{tikzpicture}
            } & \multicolumn{1}{c}{
                \begin{tikzpicture}[scale=0.40]
                    \Vertex[size=0.1, color=white]{A} 
                    \Vertex[size=0.1, color=white, x=2]{B}
                    \Edge[color=green, lw=1pt](A)(B)
                \end{tikzpicture}
            } & \multicolumn{1}{c}{
                \begin{tikzpicture}[scale=0.40]
                    \Vertex[size=0.1, color=white]{A} 
                    \Vertex[size=0.1, color=white, x=2]{B}
                    \Edge[color=red, lw=1pt](A)(B)
                \end{tikzpicture}
            } & \multicolumn{1}{c}{
                \begin{tikzpicture}[scale=0.40]
                    \Vertex[size=0.1, color=white]{A} 
                    \Vertex[size=0.1, color=white, x=2]{B}
                    \Edge[color=green, lw=1pt](A)(B)
                \end{tikzpicture}
            } & \multicolumn{1}{c}{
                \begin{tikzpicture}[scale=0.40]
                    \Vertex[size=0.1, color=white]{A} 
                    \Vertex[size=0.1, color=white, x=2]{B}
                    \Edge[color=red, lw=1pt](A)(B)
                \end{tikzpicture}
            } & \multicolumn{1}{c}{
                \begin{tikzpicture}[scale=0.40]
                    \Vertex[size=0.1, color=white]{A} 
                    \Vertex[size=0.1, color=white, x=2]{B}
                    \Edge[color=green, lw=1pt](A)(B)
                \end{tikzpicture}
            } \\[-0.5ex]
    & \multicolumn{1}{c}{1.85} & \multicolumn{1}{c}{\textbf{98.15}} & \multicolumn{1}{c}{\textbf{90.74}} & \multicolumn{1}{c}{9.26} & \multicolumn{1}{c}{\textbf{98.15}} & \multicolumn{1}{c}{1.85} & \multicolumn{1}{c}{0.00} & \multicolumn{1}{c}{\textbf{100.00}} \\
    \midrule
    \textbf{triads}\!& & & & & & & & \\[-1.5ex]
            & \multicolumn{1}{c}{
                \begin{tikzpicture}[scale=0.40]
                \Vertex[size=0.1, color=white]{A} 
                \Vertex[size=0.1, color=white, x=1, y=1]{B} \Vertex[size=0.1, color=white, x=2]{C}
                \Edge[color=green, lw=1pt](A)(B)
                \Edge[color=green, lw=1pt](A)(C)
                \Edge[color=green, lw=1pt](B)(C)
                \end{tikzpicture}
          } 
           & \multicolumn{1}{c}{
                \begin{tikzpicture}[scale=0.40]
                \Vertex[size=0.1, color=white]{A} 
                \Vertex[size=0.1, color=white, x=1, y=1]{B} \Vertex[size=0.1, color=white, x=2]{C}
                \Edge[color=red, lw=1pt](A)(B)
                \Edge[color=red, lw=1pt](A)(C)
                \Edge[color=red, lw=1pt](B)(C)
                \end{tikzpicture}
          } 
          & \multicolumn{1}{c}{
                \begin{tikzpicture}[scale=0.40]
                \Vertex[size=0.1, color=white]{A} 
                \Vertex[size=0.1, color=white, x=1, y=1]{B} \Vertex[size=0.1, color=white, x=2]{C}
                \Edge[color=red, lw=1pt](A)(B)
                \Edge[color=green, lw=1pt](A)(C)
                \Edge[color=green, lw=1pt](B)(C)
                \end{tikzpicture}
          } 
          & \multicolumn{1}{c}{
                \begin{tikzpicture}[scale=0.40]
                \Vertex[size=0.1, color=white]{A} 
                \Vertex[size=0.1, color=white, x=1, y=1]{B} \Vertex[size=0.1, color=white, x=2]{C}
                \Edge[color=green, lw=1pt](A)(B)
                \Edge[color=green, lw=1pt](A)(C)
                \Edge[color=red, lw=1pt](B)(C)
                \end{tikzpicture}
          } 
          & \multicolumn{1}{c}{
                \begin{tikzpicture}[scale=0.40]
                \Vertex[size=0.1, color=white]{A} 
                \Vertex[size=0.1, color=white, x=1, y=1]{B} \Vertex[size=0.1, color=white, x=2]{C}
                \Edge[color=green, lw=1pt](A)(B)
                \Edge[color=red, lw=1pt](A)(C)
                \Edge[color=green, lw=1pt](B)(C)
                \end{tikzpicture}
          }
          & \multicolumn{1}{c}{
                \begin{tikzpicture}[scale=0.40]
                \Vertex[size=0.1, color=white]{A} 
                \Vertex[size=0.1, color=white, x=1, y=1]{B} \Vertex[size=0.1, color=white, x=2]{C}
                \Edge[color=red, lw=1pt](A)(B)
                \Edge[color=green, lw=1pt](A)(C)
                \Edge[color=red, lw=1pt](B)(C)
                \end{tikzpicture}
          }
          & \multicolumn{1}{c}{
                \begin{tikzpicture}[scale=0.40]
                \Vertex[size=0.1, color=white]{A} 
                \Vertex[size=0.1, color=white, x=1, y=1]{B} 
                \Vertex[size=0.1, color=white, x=2]{C}
                \Edge[color=green, lw=1pt](A)(B)
                \Edge[color=red, lw=1pt](A)(C)
                \Edge[color=red, lw=1pt](B)(C)
                \end{tikzpicture}
          }
          & \multicolumn{1}{c}{
                \begin{tikzpicture}[scale=0.40]
                \Vertex[size=0.1, color=white]{A} 
                \Vertex[size=0.1, color=white, x=1, y=1]{B} \Vertex[size=0.1, color=white, x=2]{C}
                \Edge[color=red, lw=1pt](A)(B)
                \Edge[color=red, lw=1pt](A)(C)
                \Edge[color=green, lw=1pt](B)(C)
                \end{tikzpicture}
          } 
           \\[0.5ex]
    \begin{tikzpicture}[scale=0.50]       
    \Text[anchor=north east]{\texttt{\scriptsize{H}}}
    \Text[anchor=south, x=0.5, y=0.7]{\texttt{\scriptsize{G}}}
    \Text[anchor=north, x=1.3]{\texttt{\scriptsize{P}}}
    \Vertex[size=0.2, color=white]{A} 
    \Vertex[size=0.2, color=white, x=0.5, y=0.5]{B} 
    \Vertex[size=0.2, color=white, x=1]{C}
    \Edge[lw=1pt](A)(B)
    \Edge[lw=1pt](A)(C)
    \Edge[lw=1pt](B)(C)
    \end{tikzpicture} 
    & \multicolumn{1}{c}{7.41} & \multicolumn{1}{c}{1.85} & \multicolumn{1}{c}{1.85} & \multicolumn{1}{c}{33.33} & \multicolumn{1}{c}{0.00} & \multicolumn{1}{c}{\textbf{55.56}} & \multicolumn{1}{c}{0.00} & \multicolumn{1}{c}{0.00}  \\[-1.5ex]
    \begin{tikzpicture}[scale=0.50]
    \Text[anchor=north east]{\texttt{\scriptsize{R}}}
    \Text[anchor=south, x=0.5, y=0.7]{\texttt{\scriptsize{G}}}
    \Text[anchor=north, x=1.3]{\texttt{\scriptsize{P}}}
    \Vertex[size=0.2, color=white]{A} 
    \Vertex[size=0.2, color=white, x=0.5, y=0.5]{B} 
    \Vertex[size=0.2, color=white, x=1]{C}
    \Edge[lw=1pt](A)(B)
    \Edge[lw=1pt](A)(C)
    \Edge[lw=1pt](B)(C)
    \end{tikzpicture} 
    & \multicolumn{1}{c}{3.70} & \multicolumn{1}{c}{11.11} & \multicolumn{1}{c}{5.56} & \multicolumn{1}{c}{16.67} & \multicolumn{1}{c}{0.00} & \multicolumn{1}{c}{\textbf{61.11}} & \multicolumn{1}{c}{1.85} & \multicolumn{1}{c}{0.00} \\[-1.5ex]
    \begin{tikzpicture}[scale=0.50]
    \Text[anchor=north east]{\texttt{\scriptsize{R}}}
    \Text[anchor=south, x=0.5, y=0.7]{\texttt{\scriptsize{O}}}
    \Text[anchor=north, x=1.3]{\texttt{\scriptsize{A}}}
    \Vertex[size=0.2, color=white]{A} 
    \Vertex[size=0.2, color=white, x=0.5, y=0.5]{B} 
    \Vertex[size=0.2, color=white, x=1]{C}
    \Edge[lw=1pt](A)(B)
    \Edge[lw=1pt](A)(C)
    \Edge[lw=1pt](B)(C)
    \end{tikzpicture} 
    & \multicolumn{1}{c}{\textbf{59.26}} & \multicolumn{1}{c}{0.00} & \multicolumn{1}{c}{11.11} & \multicolumn{1}{c}{0.00} & \multicolumn{1}{c}{24.07} & \multicolumn{1}{c}{0.00} & \multicolumn{1}{c}{0.00} & \multicolumn{1}{c}{5.56}  \\[-1.5ex]
    \begin{tikzpicture}[scale=0.50]
   \Text[anchor=north east]{\texttt{\scriptsize{V}}}
    \Text[anchor=south, x=0.5, y=0.7]{\texttt{\scriptsize{O}}}
    \Text[anchor=north, x=1.3]{\texttt{\scriptsize{A}}}
    \Vertex[size=0.2, color=white]{A} 
    \Vertex[size=0.2, color=white, x=0.5, y=0.5]{B} 
    \Vertex[size=0.2, color=white, x=1]{C}
    \Edge[lw=1pt](A)(B)
    \Edge[lw=1pt](A)(C)
    \Edge[lw=1pt](B)(C)
    \end{tikzpicture} 
    & \multicolumn{1}{c}{5.56} & \multicolumn{1}{c}{0.00} & \multicolumn{1}{c}{29.63} & \multicolumn{1}{c}{0.00} & \multicolumn{1}{c}{18.52} & \multicolumn{1}{c}{0.00} & \multicolumn{1}{c}{0.00} & \multicolumn{1}{c}{\textbf{46.30}} \\[-1.5ex]
    \begin{tikzpicture}[scale=0.50]
     \Text[anchor=north east]{\texttt{\scriptsize{G}}}
    \Text[anchor=south, x=0.5, y=0.7]{\texttt{\scriptsize{O}}}
    \Text[anchor=north, x=1.3]{\texttt{\scriptsize{A}}}
    \Vertex[size=0.2, color=white]{A} 
    \Vertex[size=0.2, color=white, x=0.5, y=0.5]{B} 
    \Vertex[size=0.2, color=white, x=1]{C}
    \Edge[lw=1pt](A)(B)
    \Edge[lw=1pt](A)(C)
    \Edge[lw=1pt](B)(C)
    \end{tikzpicture} 
    & \multicolumn{1}{c}{3.70} & \multicolumn{1}{c}{0.00} & \multicolumn{1}{c}{20.37} & \multicolumn{1}{c}{0.00} & \multicolumn{1}{c}{16.67} & \multicolumn{1}{c}{0.00} & \multicolumn{1}{c}{0.00} & \multicolumn{1}{c}{\textbf{59.26}} \\[-1.5ex]
    \begin{tikzpicture}[scale=0.50]
   \Text[anchor=north east]{\texttt{\scriptsize{O}}}
    \Text[anchor=south, x=0.5, y=0.7]{\texttt{\scriptsize{A}}}
    \Text[anchor=north, x=1.3]{\texttt{\scriptsize{P}}}
    \Vertex[size=0.2, color=white]{A} 
    \Vertex[size=0.2, color=white, x=0.5, y=0.5]{B} 
    \Vertex[size=0.2, color=white, x=1]{C}
    \Edge[lw=1pt](A)(B)
    \Edge[lw=1pt](A)(C)
    \Edge[lw=1pt](B)(C)
    \end{tikzpicture}
    & \multicolumn{1}{c}{\textbf{59.26}} & \multicolumn{1}{c}{0.00} & \multicolumn{1}{c}{0.00} & \multicolumn{1}{c}{12.96} & \multicolumn{1}{c}{24.07} & \multicolumn{1}{c}{0.00} & \multicolumn{1}{c}{3.70} & \multicolumn{1}{c}{0.00} \\
    \bottomrule
    \end{tabular}%
  \label{tab:dyadtriad}%
  \end{scriptsize}
  \end{center}
  {\scriptsize {\bf Note}: {\linespread{0.5}\selectfont Authors' elaborations on WVS Wave 6. The table reports the percentage of the dyads and the triads that appear in at least 95\% of the 54 national cultural networks. For each structure, the table reports the percentage of times the dyad or triad appears in a certain colour-coded configuration, where green                 
  \begin{tikzpicture}[scale=0.20]
                    \Vertex[size=0.05, color=white]{A} 
                    \Vertex[size=0.05, color=white, x=2]{B}
                    \Edge[color=green, lw=1pt](A)(B)
  \end{tikzpicture}
  (red
    \begin{tikzpicture}[scale=0.20]
                    \Vertex[size=0.05, color=white]{A} 
                    \Vertex[size=0.05, color=white, x=2]{B}
                    \Edge[color=red, lw=1pt](A)(B)
  \end{tikzpicture}
  ) 
  refers to a positive (negative) partial correlation. The highest percentage is highlighted in bold. \par}}
\end{table}%
The previous analyses are based only on the edge probabilities and, thus, they ignore the sign of the dependency, i.e. whether an edge is associated to a positive or a negative partial correlation. Looking now at the edge sign will strengthen the search for commonalities and differences between the networks, particularly if the associated structure has a high weight in terms of average edge probabilities. To this end, Table \ref{tab:dyadtriad} provides the frequency of coloured-patterns for the dyads and triads for which the edge probability or the average value of their edge probabilities, respectively, is higher than the 95th percentile of the corresponding distribution
in Figure \ref{fig:nodedyadtriad}. Green and red lines refer to a positive and negative partial correlation, respectively. The results show that there is a high stability in the nature of the dependency at the level of dyads, e.g. justification of \texttt{homosexuality} and \texttt{abortion} are always positively correlated (conditionally on all the other nodes); the same is true, but in a lower percentage, in the case of \texttt{H-P}, while 
\texttt{V-M} and \texttt{G-B} are negatively correlated. The interpretation of the sign of the partial correlation depends on the directionality of the coding, as reported in Table \ref{tab:descriptives} (e.g. high level of \texttt{happiness} is positively correlated with low levels of \texttt{national pride};  the availability of people to express personal opinions by signing petitions, \texttt{voice}, is negatively correlated with progressive individual values that define \texttt{post-materialism}; the \texttt{importance of God} is negatively correlated with the prevalence of independence over obedience).
\newline
\newline
As one can notice from Table \ref{tab:dyadtriad}, there is more heterogeneity when it comes to triads. In the Table, an exemplificative subsample of triads is reported (the ones appearing at least 95\% of the time in the national cultural networks).  
\textcolor{black}{Comparing with the uniform allocation of 12.5\% in each category,  two triadic configurations prevail in the case of \texttt{H-G-P}.} They both consist of a positive correlation between \texttt{H} and \texttt{P}, as anticipated by the percentage in the first row (dyads) of Table \ref{tab:dyadtriad}, and a negative correlation between \texttt{G} and \texttt{P}, which is associated  with a negative partial correlation between \texttt{G} and \texttt{H} in 55.56\% of the national cases and  with a positive one in 33.33\% of the cases.
\textcolor{black}{However, looking at all triadic configurations reported in Table \ref{tab:dyadtriad}, while some configurations are rarely observed, there is generally a large heterogeneity in the configurations across cultural networks, pointing to significant structural differences between national cultural networks}. 
\newline
\newline
To sum up, the analysis shows how potentially different cultural networks can have cultural traits sharing a common relative position in the network or specific sub-graphs in common, while potentially similar national cultural networks can differ in the signs of the partial correlations associated to some prevalent sub-structures.

\subsection{Investigating cross-country cultural distances}
    \label{sec:distresults}

\begin{figure}[!ht]
            \begin{center}
            \caption{\small{\textit{Correlation matrix of different measures of cultural distance}. }}
            \includegraphics[width=0.65\textwidth]{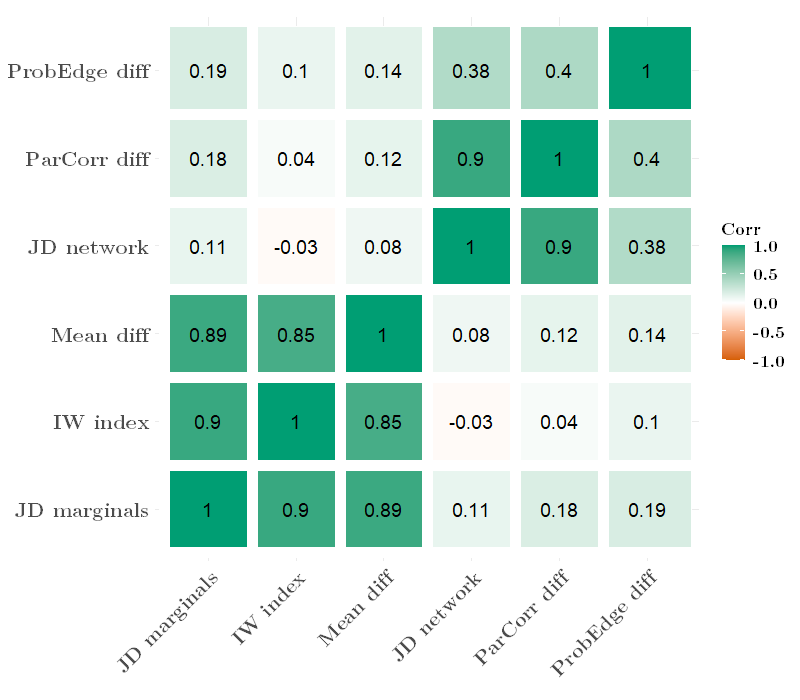}
        \label{fig:corrmap}
        \end{center}
        \vspace{-0.5cm}
{\scriptsize {\bf Note}: {\linespread{0.5}\selectfont The figure includes the correlation matrix of the different measures of cultural distance discussed in the text: \texttt{JD marginals}, \texttt{IW network}, \texttt{IW index}, the Euclidean norm of the differences in the means (\texttt{Mean diff}), the Frobenius norm of differences in the partial correlation matrices (\texttt{ParCorr diff}) and the Frobenius norm of the differences in the edge probability matrices (\texttt{ProbEdge diff}). 
\par}}
\end{figure}

\begin{figure}[!ht]
        \caption{\small{\textit{Comparison between \texttt{IW index}, \texttt{JD marginals} and \texttt{JD network} distances.}}}
        \begin{center}
            \begin{subfigure}{0.48\textwidth}
               \includegraphics[width=\linewidth]{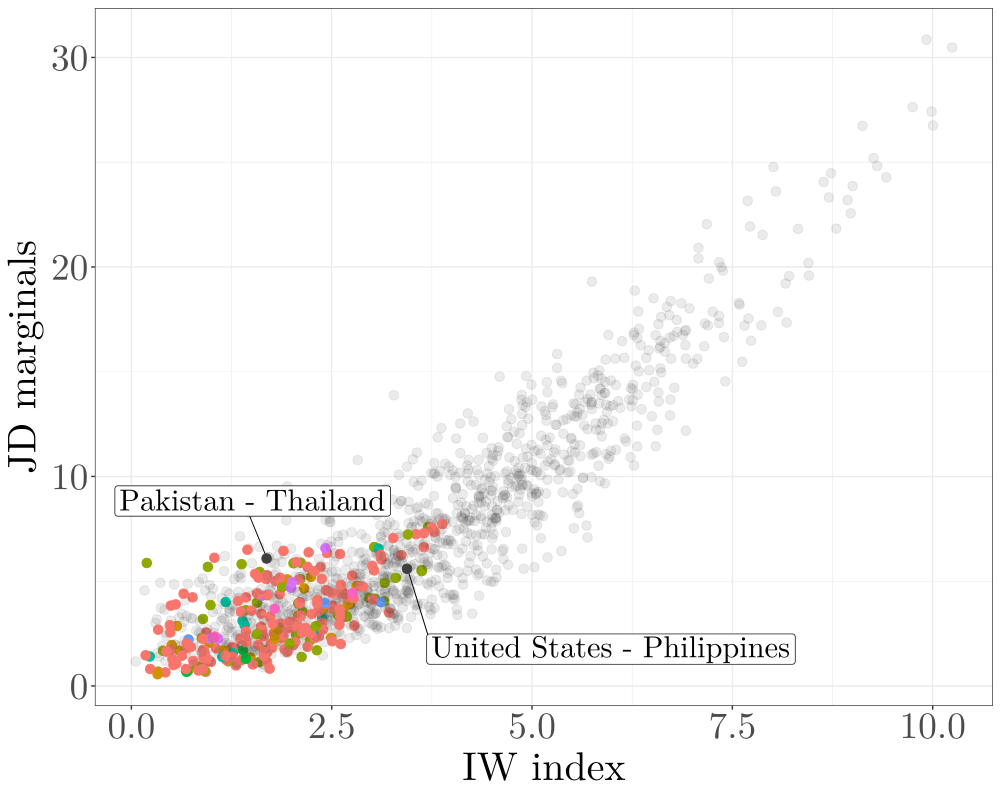} 
                \subcaption{\small{\texttt{IW index} vs \texttt{JD marginals}}}
                \label{fig:IWvsJDM}
            \end{subfigure}
            \hfill
            \begin{subfigure}{0.48\textwidth}
            \centering
               \includegraphics[width=\linewidth]{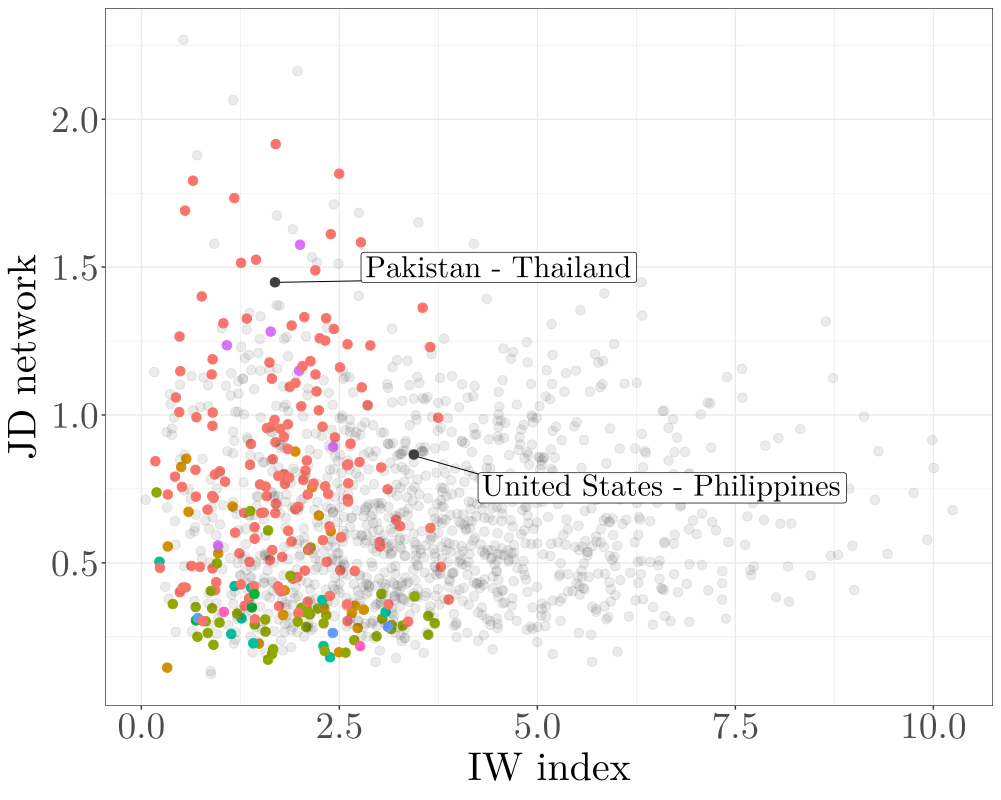} 
                \subcaption{\small{\texttt{IW index} vs \texttt{JD network}}}
                \label{fig:IWvsJDN}
            \end{subfigure}\\
            \begin{subfigure}{\textwidth}
            \centering                \includegraphics[width=0.4\linewidth]{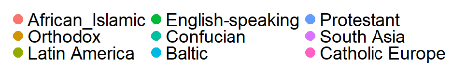} 
                \label{fig:legend}
            \end{subfigure}
        \label{fig:distances}
        \end{center}
{\scriptsize {\bf Note}: {\linespread{0.5}\selectfont Authors' elaborations on WVS Wave 6. Colours correspond to \cite{IngWel2005} groups. If a pair of countries belongs to the same IW group, then the corresponding point in the plots takes the identifying colour of that group, else the point takes the colour grey. \par}}
\end{figure}

All the elements described in the previous Sections, can now be assembled in a unifying framework that represents national cultures as networks and that quantifies the distance between national cultures as the aggregation of different components.
The \texttt{JD index} described in Section \ref{sec:distancelist} measures distance between national cultures accounting both for the marginal distribution of the individual cultural traits and for their inter-dependencies. This offers the advantage of  highlighting and explaining which specific component is influential in determining the cultural distance between countries and the relative position of a country in the cultural spectrum.  
\newline
\newline
The \texttt{JD index}, derived from Equation (\ref{eq:jeffrey}), is the Jeffreys' divergence between the corresponding inferred copula densities, while \texttt{JD marginals} and \texttt{JD network} are the two additive components of the \texttt{JD index}, corresponding to Jeffreys' divergences between the marginal distributions and between networks of the cultural traits of two countries, respectively. 
In order to compare this new proposed index of cultural distance with the original \texttt{IW index}, the latter is defined as the Euclidean distance between the countries' position on the first two dimensions of the IW Cultural Map (Figure \ref{fig:map6}). 
\newline
\newline
Figure \ref{fig:corrmap} reports the correlation between the \texttt{JD index}, split into its network and marginal component, and alternative measures of cultural distances, namely: the \texttt{IW index}; the Euclidean norm of the differences between the vector of means of the cultural traits (\texttt{Mean diff}); the Frobenius norm of the differences between the partial correlation matrices (\texttt{ParCorr diff}); and the Frobenius norm of the differences between the matrices of edge posterior probabilities (\texttt{ProbEdge diff}), for any pair of countries. The results show, firstly, a high correlation between the \texttt{IW index} and the marginal component of the \texttt{JD index} (0.9) as well as with the differences between the country means of the cultural traits (0.89). Secondly, the distances based on the marginal distributions are orthogonal to those distances that relate to the network component, as shown by the low values in the off-diagonal blocks of the correlation matrix. This is to be expected, as marginal and network measures relate to different components of the model. Indeed, measures based on the individual marginal distributions capture differences between the cultural traits at the level of their individual means, variances and higher moments, while measures based on the network component capture differences in the dependencies between the cultural traits, namely their partial correlations in the latent space, which is a feature independent from the marginal distributions. As for the \texttt{JD network} component of the \texttt{JD index}, it is interesting to note how this measure is highly correlated with differences in the partial correlation matrices (0.9), but less so with differences in the posterior edge probabilities (0.38), and thus also with differences in the presence or absence of edges in a network. This is also to be expected for a Kullback-Leibler divergence, and could motivate future work looking at alternative measures that place more emphasis on differences in the structure of the networks. 
\newline
\newline
Figure \ref{fig:distances} confirms the results from the correlation analysis, by showing a strong correlation between the \texttt{IW index} and the \texttt{JD marginals} distance (Figure \ref{fig:IWvsJDM}) but a weak correlation between the \texttt{IW index}  and the \texttt{JD network} distance (Figure \ref{fig:IWvsJDN}). This brings to a total correlation of 0.89 between the \texttt{IW index} and the \texttt{JD index}. This high value is driven, on the one hand, by the high correlation between \texttt{IW index} and \texttt{JD marginals} and, on the other hand, by the larger scale of the \texttt{JD marginals} component compared to the \texttt{JD network} component, which is the result of pairs of countries that are very different in their responses to the survey questions, captured by \texttt{Mean diff}. On the other hand, considering the proportion of the total \texttt{JD index} that is accounted for by the \texttt{JD network} component, across pairs of countries, one can notice a large heterogeneity in the weight that the network component has with respect to the total distance, with percentages ranging from 1.5\% to 58.5\%.
\newline
\newline
The added value of the encompassing new measure, both in the use of marginal distributions of the cultural traits over the aggregated means, and in the introduction of the network component, can be highlighted by inspecting more closely the overall sample of countries and some national cultural networks.
\newline
\newline
The colouring of the points in Figure \ref{fig:distances} relates to the groups defined by \cite{IngWel2005} and applied in\ Figure \ref{fig:map6} and Figure \ref{fig:nodedyadtriad}: if a pair of countries belongs to the same group, then the corresponding point in the plot takes the identifying colour of that group, else the point takes the colour grey. As it is evident, the coloured points tend to be located in the left-lower part of the plots: countries belonging to the same group tend to be culturally closer than countries belonging to different groups. This is remarkable in Figure \ref{fig:IWvsJDM}, much less so in Figure \ref{fig:IWvsJDN}. The latter shows many cases in which countries are  distant in their cultural traits dependencies (large \texttt{JD network} value) but near in their general attitude toward individual cultural traits (low \texttt{IW index} or \texttt{JD marginals} values), and vice versa. In particular, it appears that distances of African-Islamic and South Asia countries may be affected by the consideration of the cultural trait networks, as there are many instances in these groups with high \texttt{JD network} distance but low \texttt{IW index} or \texttt{JD marginals} distances (top left corner of Figures \ref{fig:IWvsJDN}). 
In terms of bilateral cultural distances, some country pairs can be very similar in the role played by one of the \texttt{JD index} components (e.g. Pakistan-Thailand and United States-Philippines have very close \texttt{JD marginals}) but they can be very distant in term of the other component. 
This heterogeneity in national cultural configurations gives strong evidence to a multifaceted global cultural diversity. 
\begin{figure}[!ht]
 \begin{center}
 \caption{The Network Structure of National Cultures}
           \begin{subfigure}{0.48\textwidth}
                \includegraphics[width=1\linewidth]{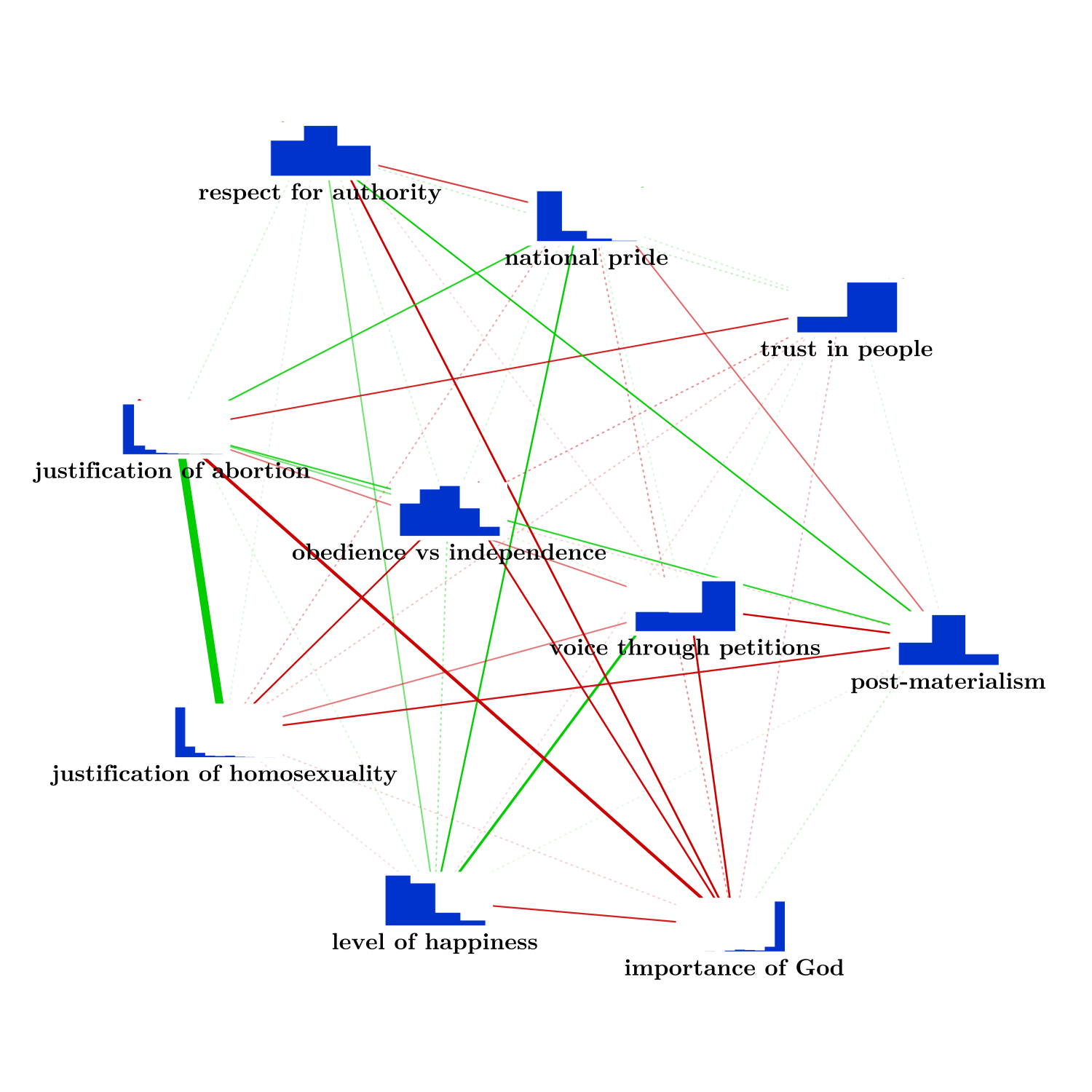} 
                \subcaption{\small{\textit{Pakistan's Cultural Network}}}
            \label{subfig:pakistan}
            \end{subfigure}
    \hfill
            \begin{subfigure}{0.48\textwidth}
                \centering
                \includegraphics[width=1\linewidth]{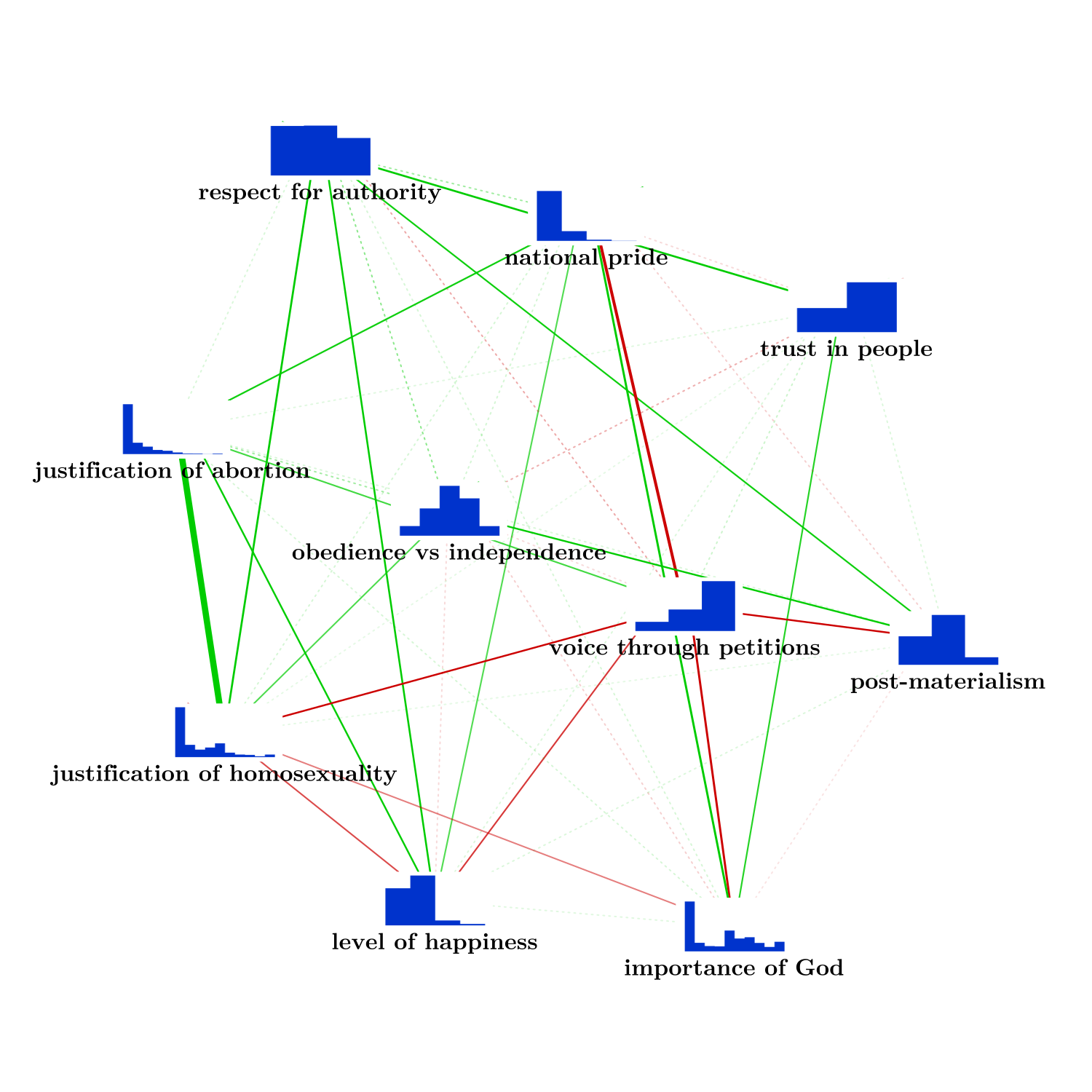}
                \subcaption{\small{\textit{Thailand's Cultural Network}}}
            \label{subfig:thailand}
            \end{subfigure}
    \hfill
            \begin{subfigure}{0.48\textwidth}
                \centering
                \includegraphics[width=1\linewidth]{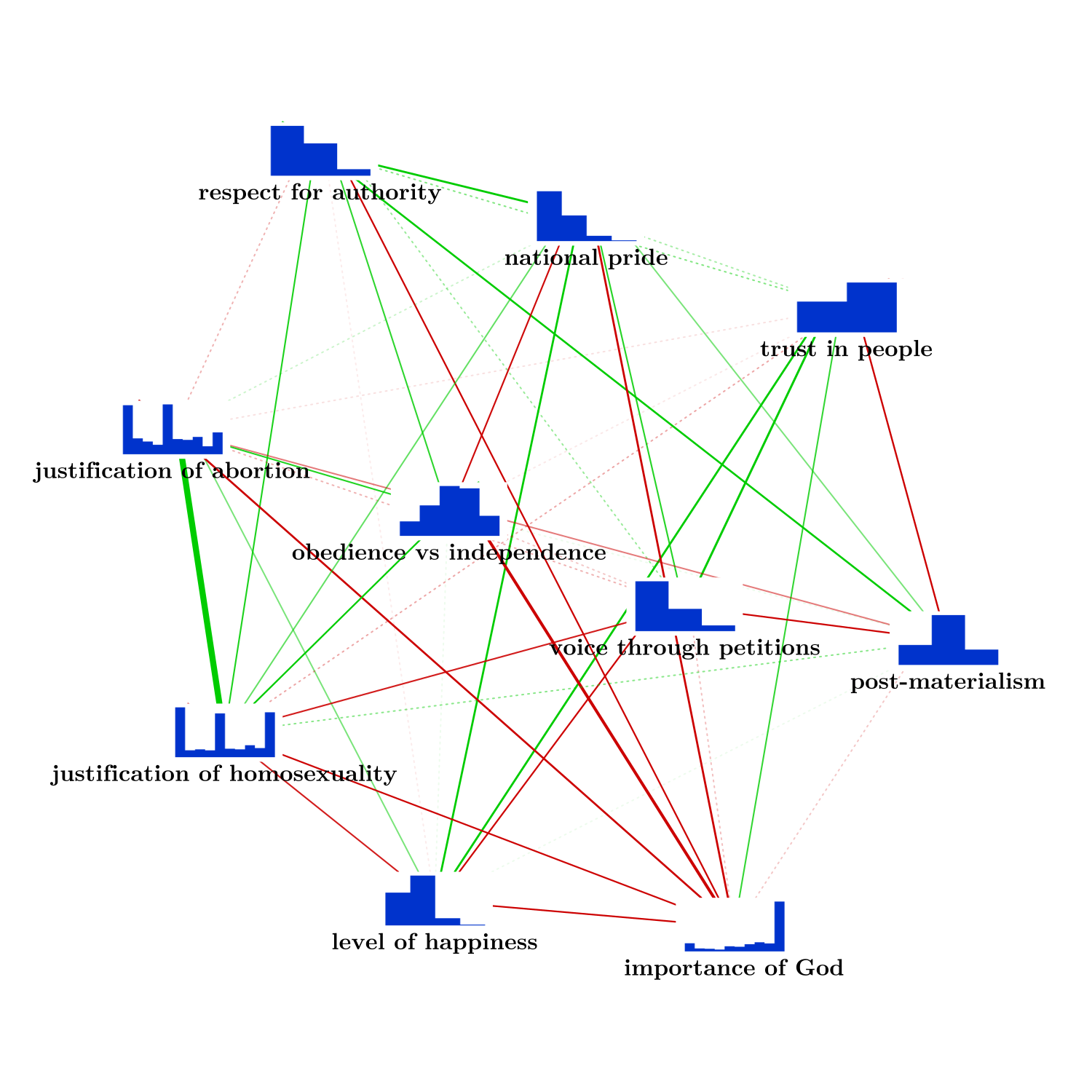}
                \subcaption{\small{\textit{United States' Cultural Network}}}
            \label{subfig:usa}
            \end{subfigure}
    \hfill
            \begin{subfigure}{0.48\textwidth}
                \centering
                \includegraphics[width=1\linewidth]{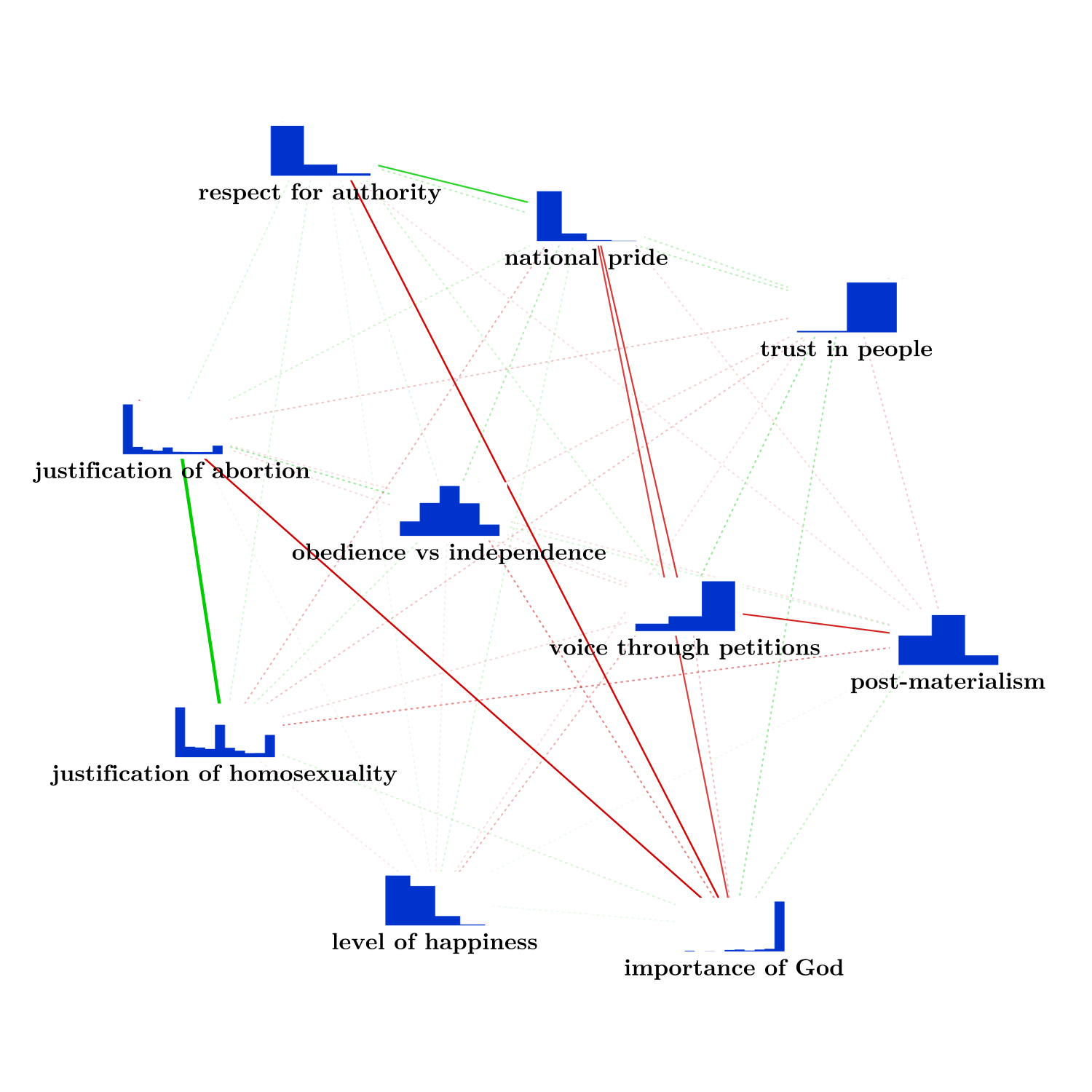}
                \subcaption{\small{\textit{{The Philippines' Cultural Network}}}}
            \label{subfig:philippines}
            \end{subfigure}
    \label{fig:nets}
     \end{center}
     \vspace{-0.3cm}
{\scriptsize {\bf Note}: {\linespread{0.5}\selectfont Authors' elaborations on WVS Wave 6. The inferred graphs for (a) Pakistan, (b) Thailand, (c) United States, and (d) The Philippines are drawn. Colouring and thickness of edges is based on the associated partial correlations (positive: green; negative: red). The line type of the edges depends on a dichotomization of the values of the posterior edge probabilities (straight line: values above 0.5; dotted line: values below 0.5), while the transparency of the edges depends directly on the posterior edge probabilities. Node labels refer to the cultural traits in Table \ref{tab:descriptives} and the node shapes are the empirical distributions of the cultural traits for each country-network.
\par}}
        \end{figure}
\newline
\newline
Some of the national cultural networks that result from the approach proposed in Section \ref{sec:gmodel}, are now taken as examples of the relevance of different elements in the emergence of cross-country cultural differences. The Pakistani, Thai, US and Filipino cases, spotted in Figure \ref{fig:distances} and already detected in Figure \ref{fig:netsummaries}, are further discussed one by one and with a comparative intent. Figure \ref{fig:nets} shows the four selected national cultural networks, and Table \ref{tab:BayesianCI} reports the credible intervals of selected network statistics of the respective country networks. Indeed, an advantage of the Bayesian procedure is the fact that one can easily go beyond point estimates, such as those used in Figures \ref{fig:netsummaries} and \ref{fig:nodedyadtriad}, by inspecting the posterior draws of graphs after convergence. Network statistics are calculated for each of these draws and the 89\% credible intervals are reported in Table \ref{tab:BayesianCI}. \textcolor{black}{The same procedure can be used to calculate credible intervals of the \texttt{JD index} itself, so as to account for the full posterior distribution of the graphs and precision matrices. }
\newline
\newline
Starting from the one of Pakistan (Figure \ref{subfig:pakistan}), the cultural network is highly wired, with many instances of high edge posterior probability. The relevance of the \texttt{importance of God} spikes out visually, but the relative homogeneity in the centrality of all nodes, as shown in Table \ref{tab:BayesianCI}, explains the low level of centrality on the Pakistani cultural network. In this respect, only node \texttt{T} turns out to be relatively disconnected from the rest of the network, emphasising the peculiar aspect of \texttt{trust} in this specific national case.
\newline
\newline
The networks of Pakistan and Thailand are very similar in terms of \texttt{density}, \texttt{centralization} and \texttt{clustering} (see Figure \ref{fig:netsummaries} and Table \ref{tab:BayesianCI}), but they are quite far apart in terms of the sign of the partial correlations between cultural traits. This is mostly due to the local structure of \texttt{importance of God} in the two countries, which tends to be negatively correlated with the other nodes in both countries but has a larger neighbourhood in the Pakistani network. This node shows also a striking difference in its marginal distribution between the two countries. In a comparative perspective, two national cultures can be very similar in terms of the overall topology of the network, but quite distant because of the specific role of one cultural trait or of a local network structure in influencing the \texttt{JD marginals} and the \texttt{JD  network} components of the \texttt{JD index}, respectively.
\newline
\newline
On the other hand, the networks of the United States and The Philippines are quite different both in terms of the skewness of the distribution of various cultural traits (e.g. \texttt{voice}) and of their global topology (see, again, Figure \ref{fig:netsummaries} and Table \ref{tab:BayesianCI}). The \texttt{density} of the United States is the highest of our sample, with a credible interval of [0.60, 0.73], as it is the level of \texttt{clustering}, with an interval of [0.56, 0.73]. This is the opposite for The Philippines, showing a \texttt{centralization} lying in the interval [0.40, 0.65]. As anticipated in Section \ref{sec:netresults}, the number of possible cultural configurations of the United States is visualised by the number of triads having a very high posterior probability for every edge (the black circles in Figure \ref{fig:nodedyadtriad}). In the case of The Philippines this is quite limited, but it is interesting to notice that the structure of the Philippine cultural network is almost a subset of the one of the United States. Another element that is worth pointing out is that countries can share some common distributional characteristics of a cultural trait (e.g. the justification of \texttt{homosexuality}) or very different ones (e.g. \texttt{voice}), having at the same time an unrelated position of those very nodes in their national cultural network (e.g. the centrality of \texttt{O} is between [6.00, 8.00] in the United States, but between [2.00, 5.00] in The Philippines).
\newline
\newline
Going back to the position of the Pakistan-Thailand and United States-Philippines country pairs in Figure \ref{fig:distances}, it is now possible to affirm that in the first case the \texttt{JD marginal} component of the cultural index is depending solely on the difference in one single cultural trait (\texttt{importance of God}), while in the United States-Philippines case, multiple cultural trait differences concur to determine the cultural distance between the two countries.
\newline
\newline
Beyond the consideration of the global topology and specific cultural traits, national cultural networks can be compared in terms of the role played by a specific dyad or triad, or by more complex cultural configurations. Here only some examples are reported.
\begin{table}[!ht]
\begin{center}
\caption{\small{\textit{Case studies: credible intervals of selected network statistics.}}}
\begin{scriptsize}
\begin{tabular}{lcccc}
\toprule
Statistics & Pakistan & Thailand & United States & The Philippines \\
\midrule
\textcolor{black}{\texttt{JD index}} & \multicolumn{2}{c}{\textcolor{black}{[1.73, 2.51]}} & \multicolumn{2}{c}{\textcolor{black}{[1.18, 1.62]}} \\
\midrule
Network &&&& \\ 
\cmidrule(l){1-1}
\texttt{density} & [0.42, 0.58] & [0.44, 0.58] & [0.60, 0.73] & [0.22, 0.38] \\
\texttt{centralization} & [0.21, 0.42] & [0.21, 0.39] & [0.13, 0.27]& [0.40, 0.65]\\
\texttt{clustering} & [0.32, 0.59] &  [0.36, 0.62] & [0.56, 0.73]& [0.10, 0.50]\\
\midrule
Node &&&& \\
\cmidrule(l){1-1}
\texttt{centrality} - \texttt{H} & [4.00, 7.00] & [4.00, 7.00] & [4.00, 6.00]& [0.00, 0.00]\\
\texttt{centrality} -  \texttt{T} & [1.00, 4.00] & [2.00, 5.00] & [4.00, 7.00]& [1.00, 5.00] \\
\texttt{centrality} -  \texttt{R} & [3.00, 6.00] & [4.00, 7.00] & [5.00, 8.00] & [2.00, 5.00]\\
\texttt{centrality} -  \texttt{V} & [4.00, 6.00] & [5.00, 7.00] & [5.00, 7.00]& [1.00, 4.00] \\
\texttt{centrality} -  \texttt{G} & [5.00, 7.00]& [3.00, 6.00]  & [6.00, 8.00]& [3.00, 6.00]\\
\texttt{centrality} -  \texttt{O} & [4.00, 7.00] & [4.00, 7.00] & [6.00, 8.00]& [2.00, 5.00]\\
\texttt{centrality} -  \texttt{A} & [4.00, 7.00] & [4.00, 7.00] & [4.00, 7.00]& [2.00, 5.00]\\
\texttt{centrality} -  \texttt{P} & [3.00, 6.00] & [4.00, 7.00] & [6.00, 9.00]& [2.00, 5.00]\\
\texttt{centrality} -  \texttt{M} & [4.00, 7.00] & [3.00, 5.00] & [3.00, 6.00]& [1.00, 4.00]\\
\texttt{centrality} -  \texttt{B} & [2.00, 5.00] & [2.00, 4.00] & [4.00, 6.00]& [1.00, 4.00]\\
\bottomrule
\end{tabular}
\end{scriptsize}
\label{tab:BayesianCI}
\end{center}
{\scriptsize {\bf Note}: {\linespread{0.5}\selectfont The Table contains the 89\% credible intervals of selected network statistics for Pakistan, Thailand, United States and The Philippines, calculated on the last 10000 posterior draws of the national cultural networks visualised in Figure \ref{fig:netsummaries}. The support of the Network statistics is [0,1], and the one of the Node \texttt{centrality} is [0, 9].
}
\par}
\end{table}
In particular, all four countries share the same prevailing partial correlation for the dyads summarised in the first row in Table \ref{tab:dyadtriad}. The difference is in the level of the posterior edge probability, which is minimal for \texttt{H-P}, in the cultural network of The Philippines, and for \texttt{G-B}, in the cultural networks of Thailand and The Philippines.  As for triads, almost all of the most common configurations of the triads reported in Table \ref{tab:dyadtriad} are present in the United States (with the exception of \texttt{R-O-A}). At the other end, only the \texttt{R-G-P} triad is present in the cultural network of The Philippines. In some cases, countries share the same triadic configuration: e.g. Pakistan and Thailand share the most common configuration of the triad \texttt{R-O-A}; in other cases countries share the same triadic configuration but with a different sign in the partial correlations: e.g. the triad \texttt{H-G-V} is present in Pakistan and the United States with high edge probabilities for all edges, but with an opposite sign in the dependency between \texttt{H-V}.
\newline
\newline
All in all, the differences and similarities between national cultures, the ones discussed in this Section but also the ones between all countries in the WVS sample, are more articulated that those captured by the \texttt{IW index}. 
\textcolor{black}{However, this does not automatically imply a higher level of dispersion of national cultures.}
\textcolor{black}{Applying a multidimensional scaling procedure \citep{kruskal1964multidimensional} to the \texttt{JD index}, and comparing it to the \texttt{IW index} through a procrustes analysis \citep{peres2001well}, it is possible to produce a new version of the Cultural Map. As shown in Figure \ref{fig:newmap6}, the new countries' coordinates are compared with those of Figure \ref{fig:map6}: the small gray dots visualise the original position of the countries according to the \texttt{IW index}, the coloured ones depict the new position derived from the \texttt{JD index}, while the Euclidean distance between the two is made evident by the gray segments.}
\begin{figure}[!h]
                \begin{center}
                \caption{\small{{The new \texttt{JD index}  Cultural Map}}}
                \includegraphics[width=\textwidth]{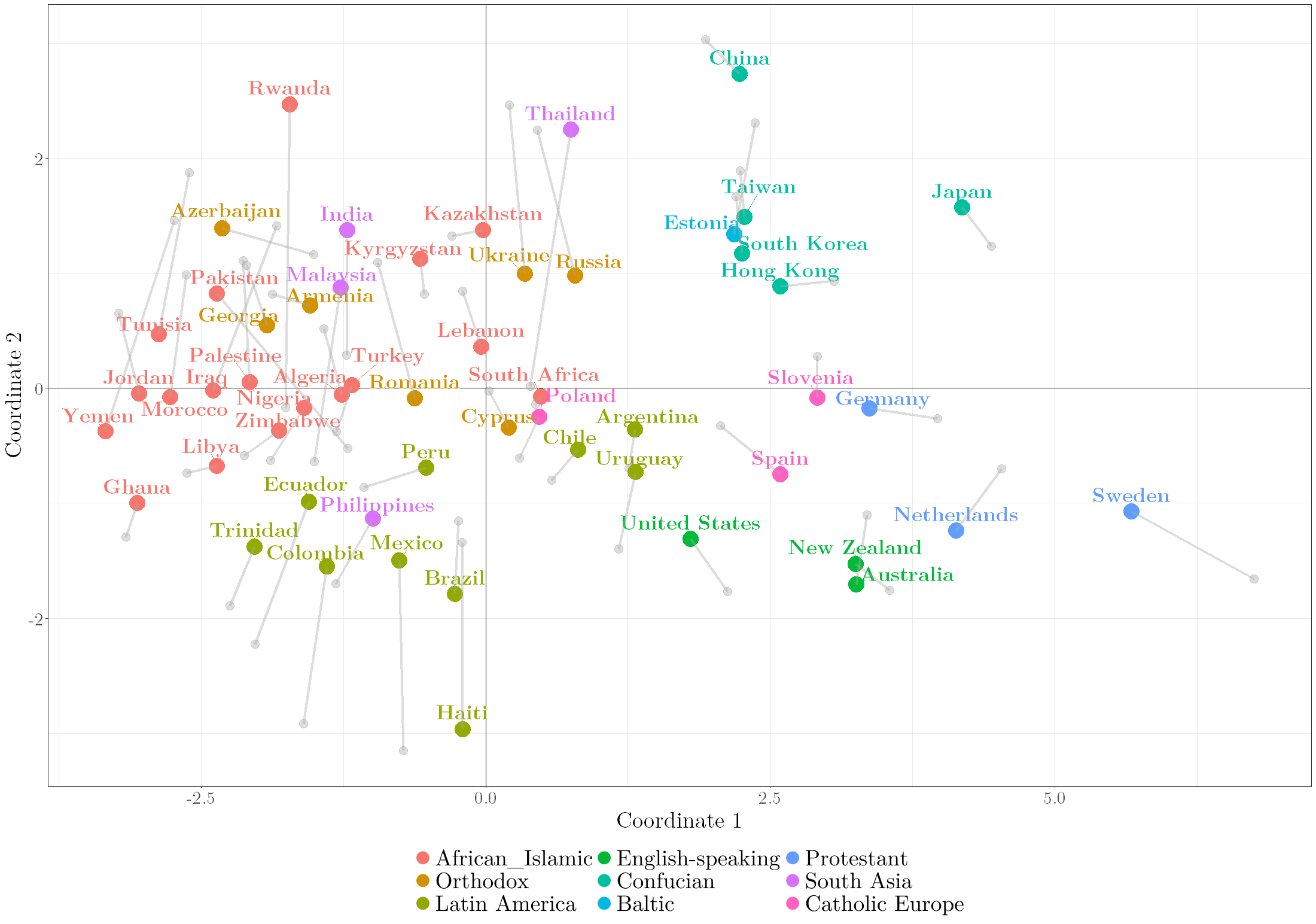}
                \label{fig:newmap6}
                \end{center}
                     \vspace{-0.1cm}
                {\scriptsize {\bf Note}: {\linespread{0.5}\selectfont Authors' elaborations on WVS Wave 6. Colors correspond to \cite{IngWel2005} groups. Gray dots correspond to the \cite{IngWel2005} index original coordinates in Figure \ref{fig:map6}. Gray segments visualize the distance between the \texttt{IW index} and the \texttt{JD index} for every country.\par}}
            \end{figure}
\textcolor{black}{An inspection of Figure \ref{fig:newmap6} reveals a slight convergence of countries coordinates towards the origin of the axes of the Cultural Map. Indeed, the average cultural distance among country-pairs is 3.08 in the case of the \texttt{JD index}, while 3.32 for the \texttt{IW index}. This in turn affects also the average within and between country groups distances, whereby the average within group distance is 1.75 in the case of the \texttt{JD index} and 1.82 for the \texttt{IW index}, while the average between country group distance is 3.36 for the  \texttt{JD index} and 3.63 for the \texttt{IW index}.  This average behaviour is, however, not shared by all countries: Thailand, India, Rwanda, Azerbaijan, Brazil, Haiti, Spain show a noteworthy distancing from the global core of the cultural space. As a result, the case study Thailand-Pakistan, that was just presented, is, for example, much closer in the \cite{IngWel2005} map (Figure \ref{fig:map6}) than in the newly derived map. This in turn leads to a significant increase of some within group distances, such as that for the South Asia countries, which increases from the 1.68 in the case of the \texttt{IW index} to the 2.24 for the \texttt{JD index}. 
Both in the case of convergence and in the one of divergence, given the high correlation between the \texttt{IW index} and \texttt{JD marginals} (Figure \ref{eq:corr}), the differences (visualized by the length of the gray segments in Figure \ref{fig:newmap6}) are mainly attributed to the \texttt{JD network} component of the \texttt{JD index}.  
}

\subsection{Identifying the factors associated to cultural distance}
\label{sec:regresults}
Having defined a unified framework for measuring culture and culture distances, this Section investigates the factors that are associated with cultural distances
. The specific form of the \texttt{JD index}, and particularly its natural decomposition into the two orthogonal components, allows to study which factors may impact on the divergence between country views about certain values or customs and/or between their inter-connectedness. To this aim, regression models are performed separately for the \texttt{JD marginals} and \texttt{JD network} component, respectively.
\newline
\newline
As cultural distances are dyadic objects, social relations models \citep{wong1982round} are adopted for this part of the analysis. These models account for statistical dependencies between distances that relate to the same node with the inclusion of random effects for each node. Coherently with the graphical modelling analysis conducted in the previous Sections, the Bayesian implementation of \cite{hoff2009multiplicative} is used, taking advantage of the \texttt{amen} R package. 
The chosen covariates for the models are split into exogenous variables, that may contribute to cultural distance either at the marginal or network level, and topological variables, that may explain differences between the structure of the cultural networks.
\newline
\newline
Exogenous variables include distances between country $m$ and country $l$ at the geographical level (variable \texttt{geographical distance} and \texttt{spatial contiguity}), similarities between countries in their historical heritage (\texttt{common empire} and \texttt{common language}), political (\texttt{polity similarity}) or economical (\texttt{gdpcc similarity}) status. These variables are traditional covariates in gravity models in economics \citep{de2011gravity, head2014gravity} and political science \citep{ward2013gravity}.
\newline
\newline
Data for most of these covariates come from the CEPII Gravity Database \citep{head2014gravity}. In particular: \texttt{geographical distance} is defined as the bilateral population-weighted distance between the most populated cities between two countries, measured in km and taken in the log scale; \texttt{spatial contiguity} is a dummy variable equal to 1 if two countries share a land border; 
\texttt{common empire} is a dummy variable equal to 1 if countries share a common coloniser post 1945; \texttt{common language} is a dummy variable equal to 1 if countries share a common official or primary language.
\newline
\newline
As for the \texttt{gdpcc similarity} and \texttt{polity similarity} covariates, the GDP per capita is extracted from the World Development Indicators World Bank collection (\url{https://www.worldbank.org/en/home}) and is measured in US dollars, Purchasing power parity in 2010; whereas polity is the ``polity2'' variable, from the  Polity IV dataset, an ordered variable that qualifies political regimes on a 21-point scale ranging from -10 (hereditary monarchy) to +10 (consolidated democracy) (\url{http://www.systemicpeace.org/inscr//p4v2015.sav},  \citep{marshall2002polity}). 
From the original values at the country level, the \citep{helpman1987imperfect} similarity index is calculated:
\begin{equation}
\texttt{X}_{ml}= 1- \left ( \dfrac{\texttt{X}_m}{\texttt{X}_m + \texttt{X}_l} \right )^2 - \left ( \dfrac{\texttt{X}_l}{\texttt{X}_m + \texttt{X}_l} \right )^2,
\label{eq:helpmanindex}
\end{equation}
with \texttt{X} indicating either GDP or polity, and $\texttt{X}_{ml}$ corresponding to  \texttt{gdpcc similarity} or \texttt{polity similarity}.
Finally, an additional variable, \texttt{IW groups}, is considered. This is a binary variable indicating whether the two countries belong to the same group used by  \cite{IngWel2005} (corresponding to the 9 groups of countries in Figures \ref{fig:map6}, \ref{fig:netsummaries} and \ref{fig:distances}). 
\newline
\newline
The results in Table \ref{tab:reg2} show how regressing \texttt{JD marginals} (models (1), (2) and (3) in Table \ref{tab:reg2}) and \texttt{JD network} (models (4), (5) and (6) in Table \ref{tab:reg2}) on a set of covariates, taking into account the dependence between dyadic observations, can shed some light into the factors associated to international cultural distances.
\newline
\newline
In model (1) the \texttt{geographical distance} between country $m$ and country $l$ \textcolor{black}{is positively correlated with the countries' cultural distance} in terms of  differences in cultural traits (\texttt{JD marginals}). In other words, geography matters in influencing the cultural proximity/distance between countries: more geographically distant countries tend to be culturally more distant. \texttt{spatial contiguity} is however not significant, \textcolor{black}{showing that, controlling for geographical distance, countries' contiguity does not accentuate or reduce the countries' cultural distance based on different cultural traits}. The inclusion of exogenous variables in model (2) does not change the qualitative message of model (1). The conditional correlation between geographical space and culture is maintained \textcolor{black}{(even though \texttt{spatial contiguity} shows some association with a reduction of cultural distance)} given that \texttt{gdpcc similarity}, \texttt{polity similarity}, and \texttt{common empire} all contribute to reduce cultural distance. Only sharing a \texttt{common language} does not seem to have a statistically significant marginal effect on cultural distance. The inclusion of \texttt{IW groups} in the set of covariates in model (3) makes the previous results more articulated: \textcolor{black}{it reduces the significance of the within-group \texttt{spatial contiguity}; it gives evidence that cultural distance can be present in groups of countries sharing the same language; and it indicates that the \texttt{JD marginals} component of cultural distance is reduced within country groups}.
\newline
\newline
\begin{table}[!ht]
\begin{center}
\caption{\small{\textit{Dyadic regression on the} \texttt{JD index} \textit{components}}}
\begin{scriptsize}
\begin{tabular}{lcccccc}
\toprule
Dependent variable $\longrightarrow$ &\multicolumn{3}{c}{\texttt{JD marginals}}  & \multicolumn{3}{c}{\texttt{JD network}} \\
\cmidrule(l){2-4}
\cmidrule(l){5-7}
Covariates    & model (1) & model (2)     & model (3)      &  model (4) & model (5) &  model (6) \\
  $\xdownarrow{0.25cm}$   &  &      &       &  & &      \\
\midrule
\multirow{2}*{intercept} & \textbf{-3.855} & \textbf{7.511} & \textbf{11.445} & \textbf{2.235}  & \textbf{2.280} & \textbf{2.630} \\[-0.5ex]
                         & [{\tiny -5.85 \ \ -1.7}]    & [{\tiny 5.45 \ \ 9.71}]  & [{\tiny 9.61 \ \ 13.45}] & [{\tiny 1.86 \ \ 2.67}]  & [{\tiny 1.87 \ \ 2.64}]  & [{\tiny 2.33 \ \ 2.95}]               \\[0.5ex]

\multirow{2}*{\texttt{geographical distance}} & \textbf{1.216} &  \textbf{1.045} & \textbf{0.517} & 0.006 & 0.015       & 0.006\\[-0.5ex]
                                              & [{\tiny 1.03 \ \ 1.4}]   &  [{\tiny 0.82 \ \ 1.22}]   & [{\tiny 0.33 \ \ 0.72}]& [{\tiny -0.01 \ \ 0.02}] & [{\tiny 0.00 \ \ 0.03}]  & [{\tiny 0.00 \ \ 0.02}]    \\[0.5ex]

\multirow{2}*{\texttt{spatial contiguity}} & -0.801 & \textcolor{black}{\textbf{-0.568}} & -0.475  & \textbf{-0.061}   & \textbf{-0.036}     & \textbf{-0.049}     \\[-0.5ex]
                                           & [{\tiny -1.52 \ \ 0.11}]  &  [{\tiny -1.31 \ \ -0.01}]   & [{\tiny -1.02 \ \ 0.14}] & [{\tiny -0.10 \ \ -0.03}] & [{\tiny -0.07 \ \ -0.01}]  & [{\tiny -0.08 \ \ -0.02}]  \\[0.5ex]
\cmidrule(l){1-2}

\multirow{2}*{\texttt{gdppc similarity}} &  & \textbf{-11.465} & \textbf{-10.689} & \textbf{-0.248} & \textbf{-0.241} & \textbf{-0.268} \\[-0.5ex]
                                         &  & [{\tiny -12.31 \ \ -10.60}]        & [{\tiny -11.63 \ \ -9.79}]&  [{\tiny -0.30 \ \ -0.20}] & [{\tiny -0.29 \ \ -0.20}]  & [{\tiny -0.31 \ \ -0.22}]               \\[0.5ex]

\multirow{2}*{\texttt{polity similarity}} & & \textbf{-12.940}  & \textbf{-11.229}    & \textbf{-0.353}       & \textbf{-0.399}       & \textbf{-0.262}                                             \\[-0.5ex]
                                         &  & [{\tiny -15.68 \ \ -10.66}]       & [{\tiny -13.41 \ \ -9.12}]& [{\tiny -0.51 \ \ -0.23}]  & [{\tiny -0.50 \ \ -0.27}] & [{\tiny -0.38 \ \ -0.15}]  \\[0.5ex]
                
\multirow{2}*{\texttt{common empire}} & & \textbf{-2.691} & \textbf{-2.757}    & \textbf{-0.085}       & \textcolor{black}{\textbf{-0.037}}        & -0.022  \\[-0.5ex]
                                     & & [{\tiny -3.17 \ \ -2.20}]       & [{\tiny -3.33 \ \ -2.32}]& [{\tiny -0.12 \ \ -0.06}]  & [{\tiny -0.06 \ \ -0.01}]     & [{\tiny -0.05 \ \ 0.00}]      \\[0.5ex]

\multirow{2}*{\texttt{common language}} & & 0.085 & \textcolor{black}{\textbf{0.434}} & \textbf{-0.046}      & \textbf{-0.035}        & \textbf{-0.034}  \\[-0.5ex]
                                        & & [{\tiny -0.30 \ \ 0.45}]       & [{\tiny 0.08 \ \ 0.78}]& [{\tiny -0.07 \ \ -0.02}]  & [{\tiny -0.06 \ \ -0.02}]      & [{\tiny -0.06 \ \ -0.02}]       \\[0.5ex]

\cmidrule(l){1-3}

\multirow{2}*{\texttt{IW groups}} & &  & \textbf{-2.060} &  \textcolor{black}{-0.026}    & 0.000      & 0.000                   \\[-0.5ex]
                                 & & & [{\tiny -2.39 \ \ -1.71}]       & [{\tiny -0.04 \ \  0.00}] & [{\tiny -0.02 \ \ 0.02}]  & [{\tiny -0.02 \ \ 0.02}]    \\[0.5ex]

\cmidrule(l){1-4}

\multirow{2}*{\texttt{density}} & & & & \textbf{-5.764} & \textbf{-9.223}       & \textbf{-11.058}             \\[-0.5ex]
                                & & & & [{\tiny -7.48 \ \ -4.09}] & [{\tiny -10.69 \ \ -7.59}]   & [{\tiny -12.55 \ \ -9.69}]            \\[0.5ex]
                
\multirow{2}*{\texttt{centralization}} & & & & 0.130         & \textcolor{black}{-0.018} & -0.102                    \\[-0.5ex]
                                       & & & & [{\tiny -0.28 \ \ 0.55}] & [{\tiny -0.41 \ \ 0.32}]    & [{\tiny -0.42 \ \ 0.23}]           \\[0.5ex]

\multirow{2}*{\texttt{clustering}} & & & & \textbf{2.868}         & \textbf{6.019}  & \textbf{6.849}                  \\[-0.5ex]
                                   & & & & [{\tiny 1.63 \ \ 4.49}] &   [{\tiny 4.69 \ \ 7.18}]     &   [{\tiny 5.73 \ \ 8.05}]        \\[0.5ex]

\cmidrule(l){1-1} 
\cmidrule(l){5-5} 

\multirow{2}*{\texttt{H}---\texttt{P}} & & & & & 0.084                              &           \\[-0.5ex]
                                       & & & & & [{\tiny -0.24 \ \ 0.46}]            &     \\[0.5ex]
                
\multirow{2}*{\texttt{V}---\texttt{M}} & & & & & 0.126   &       \\[-0.5ex]
                                       & & & & & [{\tiny -0.17 \ \ 0.52}]  &  \\[0.5ex]

\multirow{2}*{\texttt{G}---\texttt{B}} & & & & & \textbf{1.285}     &     \\[-0.5ex]
                                       & & & & & [{\tiny 1.03 \ \ 1.59}]   & \\[0.5ex]
                
\multirow{2}*{\texttt{O}---\texttt{A}} & & & & & \textbf{2.321}    &      \\[-0.5ex]
                                       & & & & & [{\tiny 2.18 \ \ 2.48}] &   \\[0.5ex]

\multirow{2}*{\texttt{H}--\texttt{G}--\texttt{P}} & & & & & & \textbf{0.279}          \\[-0.5ex]
                                                  & & & & & & [{\tiny 0.18 \ \ 0.37}]    \\[0.5ex]

\multirow{2}*{\texttt{R}--\texttt{G}--\texttt{P}} & & & & & & \textbf{0.357}         \\[-0.5ex]
                                                  & & & & & & [{\tiny 0.27 \ \ 0.43}]    \\[0.5ex]
                
\multirow{2}*{\texttt{R}--\texttt{O}--\texttt{A}} & & & & & & \textbf{0.323}          \\[-0.5ex]
                                                  & & & & & & [{\tiny 0.19 \ \ 0.44}]    \\[0.5ex]

\multirow{2}*{\texttt{V}--\texttt{O}--\texttt{A}} & & & & & & \textbf{0.543}          \\[-0.5ex]
                                                  & & & & & & [{\tiny 0.42 \ \ 0.65}]    \\[0.5ex]
                
\multirow{2}*{\texttt{G}--\texttt{O}--\texttt{A}}  & & & & & & \textbf{0.787}          \\[-0.5ex]
                                                   & & & & & & [{\tiny 0.72 \ \ 0.86}]    \\[0.5ex]

\multirow{2}*{\texttt{O}--\texttt{A}--\texttt{P}} & & & & & & \textcolor{black}{\textbf{0.267}}          \\[-0.5ex]
                                                  & & & & & & [{\tiny 0.12 \ \ 0.38}]    \\[0.5ex]
\midrule
\midrule
countries & 51 &  51 & 51 & 51 & 51 & 51 \\
$\sigma^2_{\alpha}$   & 5.868 & 4.571 & 4.033 & 0.060 & 0.056 & 0.055 \\
$\sigma^2_{\epsilon}$ & 6.128 & 4.942 & 4.768 & 0.015 & 0.012 & 0.011 \\
\bottomrule
\end{tabular}
\end{scriptsize}
\label{tab:reg2}
\end{center}
{\scriptsize {\bf Note}: {\linespread{0.5}\selectfont The regression models are fitted using a mixed effect model (through the \texttt{ame} function in the \texttt{amen} R package
\citep{hoff2009multiplicative}). The analysis is conducted on the \textcolor{black}{51 countries (i.e. the 54 countries, excluding Taiwan, Palestine, and Hong Kong)} for which all exogenous variables are available. After 10000 MCMC iterations, the Table reports the estimated regression coefficients, their 89\% highest density intervals in square brackets (\texttt{ci} function in the \texttt{bayestestR} R package), the variance of the random effects ($\sigma^2_{\alpha}$) and the variance of the errors ($\sigma^2_{\epsilon}$). The bold font indicates the values that are significantly different from zero. 
}
\par}
\end{table}
Looking now at the \texttt{JD network} component of the cultural distance, it is of interest to investigate further whether specific differences in the structure of the cultural networks may be influential in pulling two countries closer or in pushing them apart in the cultural spectrum. The analyses in Sections \ref{sec:netresults} and \ref{sec:distresults} point to the consideration of covariates representing summary statistics at the network level, namely \texttt{density}, \texttt{centralization} and \texttt{clustering} (model (4) in Table \ref{tab:reg2}), as well as of covariates capturing the weight associated to specific dyads and triads in the network (models (5) and (6), respectively, in Table \ref{tab:reg2}). Network statistics are inserted in the regression model as dyadic variables using the same Helpman index of equation (\ref{eq:helpmanindex}) but with \texttt{X}$_m$ denoting now the network statistics for country $m$.
In contrast to this, dyads/triads are inserted in the model as Frobenius norm of the differences between the associated sub-matrices of the partial correlation matrices of the two countries, standardised by the Frobenius norm of the difference between the two full partial correlation matrices. Here, large values imply distances between countries in terms of the network components, accounting for the strength and sign of associations. Given the high correlation between the distance calculated for a triad and the same measure relative to dyads included in the given triad, the analysis considers two separate models, one with the addition of dyads only (model (5) in Table \ref{tab:reg2}) and one for triads only (model (6) in Table \ref{tab:reg2}). The model with just the three network statistics being added is also considered (model (4) in Table \ref{tab:reg2}). The results in Table \ref{tab:reg2} show how a similarity between the densities of two networks contributes to their cultural proximity (negative coefficient in all three models), while, at a given density, the similarity in the clustering coefficient tends to be positively correlated with the network cultural distance. \textcolor{black}{Moreover, the inclusion of dyadic and triadic sub-structures makes \texttt{geographical distance} and the \texttt{IW groups} insignificant, showing a lack of association of this variable with the \texttt{JD network} component of cultural distances.} Similarly, the results from model (5) show how differences in the partial correlations corresponding to the edges \texttt{G-B} and \texttt{O-A} are associated to an increased cultural distance between countries.  This is indeed what is observed also in the case studies in Figure \ref{fig:nets}, such as, for example, the different relevance in the \texttt{G-B} dyad and in the \texttt{V-O-A} and \texttt{G-O-A} triads, confirmed in model (6).
\newline
\newline
Finally, it is interesting to notice how some of the covariates, such as \texttt{gdpcc similarity} and \texttt{polity similarity}, are correlated significantly with both \texttt{JD marginals} and \texttt{JD network}, while others contribute to cultural distance only through one component of the \texttt{JD index}. For example, \texttt{spatial contiguity} is not associated to cultural traits of adjacent countries being similar to each other, but has an influence on the similarity of the topology of the two cultural networks. Furthermore, sharing a common past  (\texttt{common empire}) or being in the same \texttt{IW group} is significantly correlated with \texttt{JD marginals} but has \textcolor{black}{little or no statistical effect (once controlling for relevant dyads or triads) on the \texttt{JD network component}. Instead, sharing a \texttt{common language} reduces the divergence in the network structure of national cultures, and thus appears more correlated with the interdependence of the cultural traits than with the similarity in the cultural traits per sè.}

\section{Conclusions}
\label{conclusions}
The availability of WVS data, with a large number of countries now taking part in the survey, has provided the ground for the quantitative study in this paper, aimed at developing a unifying framework for measuring national cultures and cross-country cultural distances and for identifying their contributing factors. Taking the \citep{IngWel2005} Cultural Map as a benchmark, this paper extends the underlying methodology with the use of graphical modelling approaches. Due to the discrete nature of survey data, the paper opts for Gaussian copula graphical models and adopts the latest Bayesian procedure for inference. In deriving a new  \emph{network index of cultural distance} through this methodology, three main innovations are of notice. Firstly, the cultural traits for each country are described by their marginal distributions, empirically estimated from the data, and distances between countries are measured by distances between these distributions, rather than between aggregated measures, such as mean or PCA contributions. Secondly, graphical models provide a way of inferring the inter-connectedness between cultural traits of individual countries, which is not considered in existing measures of cultural distance. Thirdly, the proposed network index of cultural distance, defined as \textcolor{black}{an approximation of} the Jeffreys' divergence between the corresponding Gaussian copula graphical models, combines naturally  the two orthogonal components of the cultural distance, that between the marginal distributions, \textcolor{black}{itself decomposed into the individual cultural traits}, and that between the cultural networks. A clear message of this paper, supported by the results, is that the networks of cultural traits are a valuable information in defining national cultures and in determining the distance between countries in the cultural spectrum. \textcolor{black}{Compared to previous methods, a more articulated explanation of the cultural distance between two countries can be given, where  its key contributing factors can be more easily identified.} 
\newline
\newline
\textcolor{black}{The approach developed in this paper can be adapted easily to different data, such as those coming from different waves of the WVS or alternative surveys, such as the ones in the tradition of \cite{hofstede1984culture} or \cite{schwartz1994beyond}, or the one from the GLOBE study \citep{house2004culture}. Similarly, it can  be adapted easily to a different choice of cultural traits, set of countries or even sub-national spatial units, making it possible to empirically analyse the cultural homogeneity of a country or the coexistence of regional or ethnic cultures inside the national boundaries.} A hidden advantage of Bayesian methods is that they naturally account for missing data, which arise often in survey data collection. The only constraint for the analysis is that the same selection of cultural traits \textcolor{black}{(survey questions with ordered responses) must be common to all countries included in the analysis, so that meaningful comparisons can be made between the resulting networks. Finally, different grouping of countries can be used at the validation stage or endogenously determined within the analysis.}
\newline
\newline
\textcolor{black}{As shown in the last part of the paper, associations between cultural distance and other distances between countries, such as geographical, political or economical, are often of interest. Various studies have been conducted in relation to this, particularly in the economic literature (e.g. see \cite{spolaore2018ancestry} for a recent contribution). Future research can look at the implications of the new network-based measure of cultural distance in the context of these studies. } 
\newline
\newline
\textcolor{black}{Finally, while concentrating on the cross-sectional dimension of culture
, this paper has disregarded the dynamic dimension of the phenomenon, and relevant questions related to the origin of national culture, the persistence of it, and its evolution in time. Notwithstanding the difficulty of dealing with non-homogeneous data across the different waves of the WVS, the temporal evolution of culture is a matter of great interest and can be modelled with the use of dynamic network models. Future work could also explore the possible relation between specific topological configurations  of the network in the past (a.k.a. specific sub-networks) and the evolution of the cultural network itself.
}
\newline
\newline
\textcolor{black}{The data and the code for calculating the new cultural distance and for reproducing the analyses in this paper is available at \url{https://github.com/RRondinelli/Cultural-Networks}. }
\clearpage

\bibliographystyle{apalike}
\bibliography{references}

\end{document}